\documentclass[aps,prb,twocolumn,superscriptaddress,showpacs, final]{revtex4-2} 
\usepackage[normalem]{ulem}
\usepackage{amssymb}
\usepackage{suffix}
\usepackage{mathtools}
\usepackage[utf8]{inputenc}
\usepackage{booktabs}
\usepackage{cases}
\usepackage[multiple]{footmisc}
\usepackage{dcolumn}
\usepackage{color,soul}
\usepackage{rotating}
\usepackage{perpage}
\usepackage{xcolor}
\usepackage{soul}
\usepackage{tikz}
\usepackage[T1]{fontenc}
\usepackage{etoolbox}
\usepackage{graphics}
\usepackage{graphicx}
\usepackage{subcaption}

\usepackage{siunitx}
\usepackage{hyperref}
\usepackage{float}
\usepackage{multirow}
\usepackage{mathtools}
\usepackage{bm}
\usepackage{url}
\usepackage{tikz}
\usepackage{tikz-3dplot}
\usepackage [outdir=./]{epstopdf}
\usepackage{bbold}
\usepackage{changes}

\allowdisplaybreaks
\begin{document}
\title{Floquet Topological Phases and Anomalous Hall Signatures in Irradiated Two-dimensional $d_{xy}$-Wave Altermagnets}

\author{Hosein Cheraghchi}
\email{cheraghchi@du.ac.ir}
\address{School of Physics, Damghan University, P.O. Box 36716-41167, Damghan, Iran}

\date{\today}
\vspace{1cm}
\newbox\absbox
\begin{abstract}
We study Floquet topological phases in two-dimensional $d_{xy}$-wave altermagnets driven by off-resonant circularly polarized light and subject to an out-of-plane magnetization induced via extrinsic exchange coupling from a proximate ferromagnet. Using a lattice Floquet formulation, we show that the system is governed by a driving parameter $\beta$ that controls the emergence of distinct gap-closing points and associated topological phases. For $|\beta|>1$, topology is dominated by anisotropic Dirac points at high symmetry points, leading to Chern phases with $|\mathcal{C}|=2$. For $|\beta|<1$, light-induced off-symmetry $G$ points appear in four families in the Brillouine zone, enabling higher Chern phases up to $|\mathcal{C}|=4$.

Low-energy analysis reveals that high symmetry points host anisotropic massive Dirac fermions, while $G$ points realize generalized two-dimensional anisotropic Dirac points with fully momentum-dependent pseudospin structure, leading to distinct Berry curvature distributions. In the metallic regime, the anomalous Hall conductivity provides an experimental signature of these Floquet topological phases, exhibiting sharp features associated with Berry curvature accumulation near local gap-closing regions.
\end{abstract}

\maketitle

\section{Introduction}
\begin{figure}
\includegraphics[width=\linewidth]{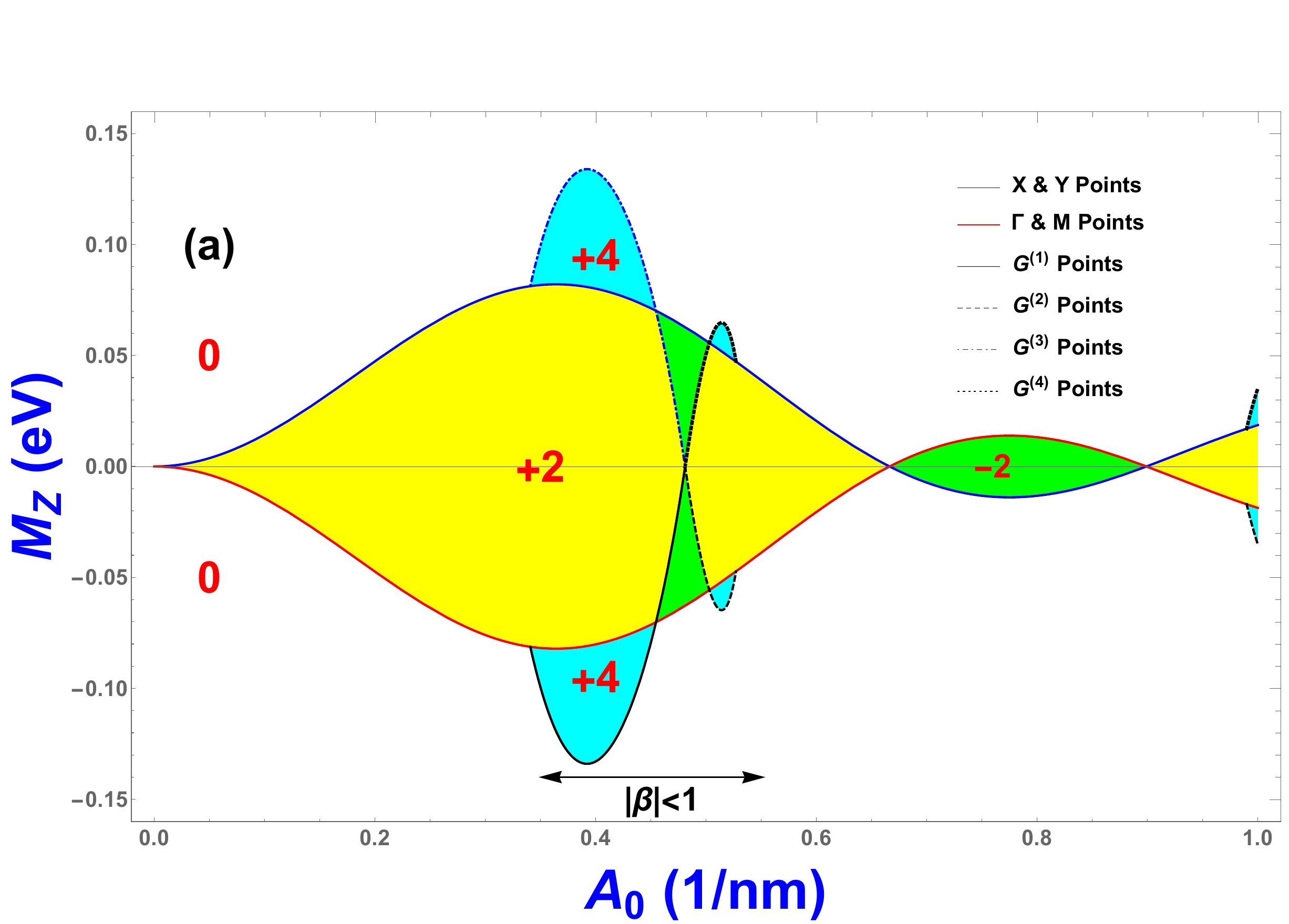}
\includegraphics[width= \linewidth]{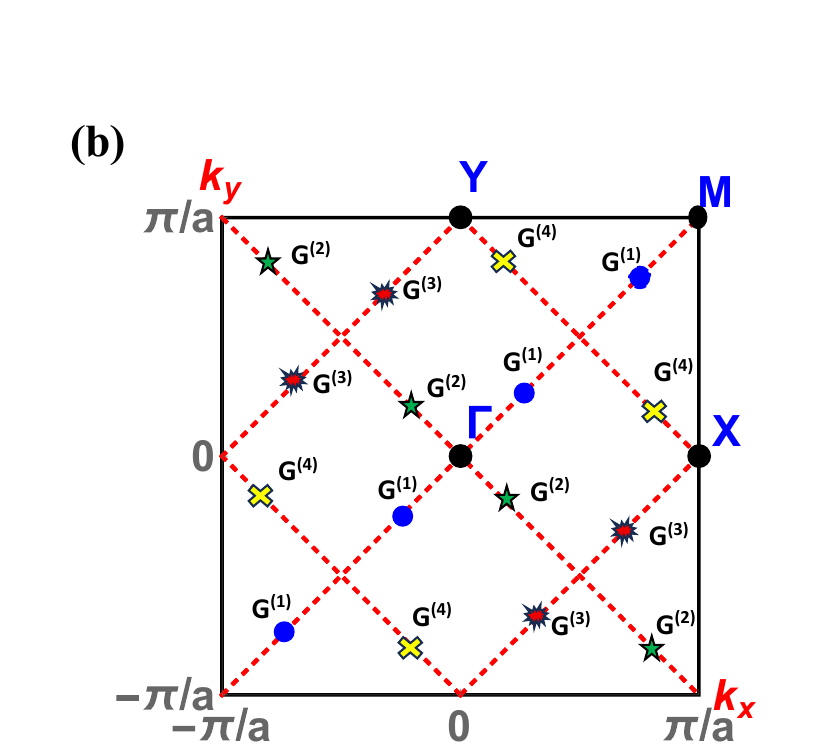}
\caption{(a) The phase diagram of $d_{xy}$- wave altermagnet irradiated by right-handed circularly polarized light and in the presence of out-of-plane magnetization. The gap-closing boundaries are determined by the effective Floquet Hamiltonian obtained in a high frequency regime. (b) the gap closing points is departed into two families; high-symmetry points, light-induced off-symmetry Dirac points. There are four members in the G-family; $G^{(1)} ... G^{(4)}$. Each member of G family contains four Dirac point distributed among the Brillouine zone based on information given in Table.~\ref{tab:dirac}. The parameters are $\lambda=0.35 ~\text{eV}$ and the altermagnetic spin splitting, $J=1.0 ~\text{eV}$. The light frequency is considered to be $\Omega=2~\text{eV}$.
 }
\label{phase_xy}
\end{figure}
Altermagnets are a recently established class of collinear compensated magnets that combine vanishing net magnetization with symmetry-protected momentum-dependent spin splitting \cite{Smejkal,PRX22,Nat_review,alter}. The alternating spin polarization originates from crystal symmetries relating opposite-spin sublattices and gives rise to characteristic $d$-, $g$-, and $i$-wave spin textures \cite{Roig2024}. In the presence of spin-orbit coupling (SOC), such momentum-space spin splitting generates finite Berry curvature and intrinsic anomalous Hall responses despite the absence of a macroscopic magnetic moment, making altermagnets a promising platform for topological phases and spintronic applications \cite{RuO1,RuO2,QAHE1,MoO,RbCrS,tunableQAHE}.

Periodic driving provides a powerful route for engineering topological band structures beyond equilibrium \cite{oka2009,kitagawa,rudner,rudner_review,Bukov,eckart}. Recent studies have shown that Floquet driving can induce higher-order SOC, persistent spin textures, spin-polarization control, odd-parity altermagnetic states, anomalous Hall responses, and topological phase transitions in altermagnets \cite{Ghorashi,Zou2025,Yarmohammadi2026,Tian2026,Ganguli2025,Hoi2026}. While several of these works rely on low-energy descriptions around selected high-symmetry points \cite{Yarmohammadi2026,Ghorashi,Hoi2026}, others have explored Floquet topology and Hall responses using lattice models \cite{Ganguli2025,Zou2025}. Nevertheless, the role of light-induced SOC acting throughout the entire Brillouin zone giving rise to the global topology and generating unconventional off-symmetry topological critical points remain largely unexplored.

In most two-dimensional topological systems, phase transitions are governed by Dirac points located at high-symmetry momenta, whose topology is determined by chirality and mass inversion \cite{haldane,Geometrical}. A lattice-based Floquet description can reveal a richer scenario in which light-induced SOC acts throughout the entire Brillouin zone and generates additional off-symmetry band-touching points\cite{Cheraghchi2025}. Unlike the anisotropic Dirac fermions that may emerge at high-symmetry momenta, these off-symmetry critical points need not belong to the conventional massive-Dirac universality class and can exhibit distinct low-energy structures and Berry-curvature distributions. Understanding the topology associated with such unconventional critical points is essential for identifying new mechanisms for high-Chern-number phases and anomalous Hall responses \cite{KMnBi}.

Importantly, the Floquet phases considered here are not restricted to fully gapped insulating states. In metallic regimes, topology is encoded in Berry-curvature hotspots concentrated near avoided crossings generated by the periodic drive. Although the Hall response is generally not quantized, Haldane's Fermi-surface formulation shows that the intrinsic anomalous Hall conductivity (AHC) remains directly sensitive to the geometric structure of electronic states near the Fermi level \cite{Haldane_2004}. Consequently, pronounced variations and sign reversals of the AHC provide experimentally accessible signatures of driven Berry-curvature-induced topology.

In this work, we investigate a two-dimensional $d_{xy}$-wave altermagnet subjected to off-resonant circularly polarized light (CPL) and an out-of-plane exchange field. Building on recent studies of Floquet-induced topology in altermagnets \cite{Zou2025,Cheraghchi2025}, we show that the Floquet system is governed by a dimensionless light-induced parameter $\beta$ separating two distinct topological regimes, where $\beta$ measures the relative strength of the renormalized intrinsic SOC and the Floquet-induced higher-harmonic SOC. For $|\beta|>1$, topology is controlled by anisotropic Dirac points located at time-reversal-invariant momenta (TRIM), yielding phases with $|C|=2$. In contrast, for $|\beta|<1$, Floquet driving generates four families of off-symmetry gap-closing points distributed throughout the Brillouin zone, enabling higher-Chern-number phases up to $|C|=4$ \cite{KMnBi}.Unlike previous low-energy Floquet treatments and recent lattice-based studies of Floquet altermagnets  \cite{Ganguli2025,Zou2025}, our lattice-based formulation incorporates Floquet-engineered SOC throughout the entire Brillouin zone\cite{Dabiri2022,Cheraghchi2025}. This global modification of band geometry gives rise to unconventional off-symmetry topology and leads to pronounced signatures in the AHC, including strong variations and sign reversals even in metallic regimes due to Berry-curvature hotspots near locally gapped crossings.

The paper is organized as follows. In Sec.~\ref{S2}, we introduce the static lattice model and derive its real-space tight-binding representation. Section~\ref{S3} presents the Floquet formalism and the analytical derivation of the effective high-frequency Hamiltonian. In Sec.~\ref{S4}, we construct the topological phase diagram as a function of the light intensity and the out-of-plane magnetization. We further analyze the associated topological phase transitions by evaluating the Chern number contributions of the gap-closing points and investigating the evolution of the Berry curvature across the phase boundaries. Section~\ref{S5} is devoted to the low-energy Floquet theory around the gap-closing points, where we characterize the anisotropy of the Dirac cones and elucidate the nature of the unconventional off-symmetry critical points. We also demonstrate how the resulting topological phases manifest themselves through signatures in the intrinsic AHC. Finally, the main conclusions are summarized in Sec.~\ref{S6}.

\section{Static Lattice Model}\label{S2}
The effective two-band Hamiltonian for a $d_{xy}$-wave altermagnet on a square lattice can be represented as the following 
\begin{equation}
\begin{aligned} 
H_{\text{static}}(k_x,k_y)&=2 t [2-\cos{(k_x a)}-\cos{(k_y a)}] \sigma_0\\
&+2 J \sin{(k_x a)}\sin{(k_y a)}\sigma_z\\
&+ \lambda[\sin{(k_x a)}\sigma_y-\sin{(k_y a)}\sigma_x]\\
&+M_z \sigma_z
\label{hamilk} 
\end{aligned} 
\end{equation}

Here, $\sigma_i$ $(i=x,y,z)$ are the Pauli matrices in spin space, and $\sigma_0$ denotes the identity matrix. The first term is represented the kinetic energy written for a square lattice with lattice parameter $a=5$~\text{nm}. The intrinsic altermagnetic exchange field is represented in the second term which produces a nonrelativistic momentum-dependent spin splitting with alternating sign. The relativistic Rashba spin–orbit coupling (RSOC) corresponds to the third term and a uniform external out-of-plane magnetization is also assumed in the last term. This uniform Zeeman-like spin splitting can be created by means of different sources. The sublattice magnetization or induced out-of-plane magnetization in altermagets (with zero net intrinsic magnetization) can be created by applying an external Zeeman magnetic field, extrinsic exchange couplings originated from neighboring to a ferromagnet, or by using the Edelstein effect due to the spin splitting under an in-plane electric field. The low-energy effective Hamiltonian is wel-known and is achievable by expanding Hamiltonian given in Eq.~\ref{hamilk} around the $\Gamma$ point. The low-energy Hamiltonian for the $d_{xy}$-wave symmetry reads as
\begin{equation}
\begin{aligned} 
\lim_{\bf{k} \to 0} H^{\text{static}}(\bf{k})&= t a^2[k_x^2+k_y^2]\sigma_0+J a^2 [k_xk_y]\sigma_z\\
&+\lambda a [k_x \sigma_y-k_y \sigma_x]+M_z \sigma_z
\label{eq:hamilchiral} 
\end{aligned} 
\end{equation}


\begin{table*}[t]
\centering
\caption{Gap-closing conditions and chiralities for light-induced Dirac points. 
The parameter $\beta = \frac{\mathcal{J}_0\lambda}{2\lambda'}$ satisfies $|\beta|<1$ for these points to exist.}
\begin{tabular*}{\linewidth}{@{\extracolsep{\fill}} c c c c c @{}}
\toprule
Point & Location condition & $\beta$ range & Gap‑closing $M^{\text{Closing}}_z =$ & Chirality $j_z$ \\
\midrule
$G^{(1)}$ & $k_y= +k_x,\;\cos^2(k_x a)=\beta$ & $[0,1]$ & $-\mathcal{J}_0 J(1-\beta)-J'\beta$ & $+(4\lambda'a)^2\beta^2(1-\beta)>0$ \\
$G^{(2)}$ & $k_y=- k_x,\;\cos^2(k_x a)=-\beta$ & $[-1,0]$ & $+\mathcal{J}_0 J(1+\beta)+J'\beta$ & $+(4\lambda'a)^2\beta^2(1+\beta)>0$ \\
$G^{(3)}$ & $k_y=\pm\frac{\pi}{a}+k_x,\;\cos^2(k_x a)=\beta$ & $[0,1]$ & $+\mathcal{J}_0 J(1-\beta)+J'\beta$ & $-(4\lambda'a)^2\beta^2(1-\beta)<0$ \\
$G^{(4)}$ & $k_y=\pm\frac{\pi}{a}-k_x,\;\cos^2(k_x a)=-\beta$ & $[-1,0]$ & $-\mathcal{J}_0 J(1+\beta)-J'\beta$ & $-(4\lambda'a)^2\beta^2(1+\beta)<0$ \\
\bottomrule
\end{tabular*}
\label{tab:dirac}
\end{table*}

\begin{figure}
\centering
\includegraphics[width=0.7\linewidth]{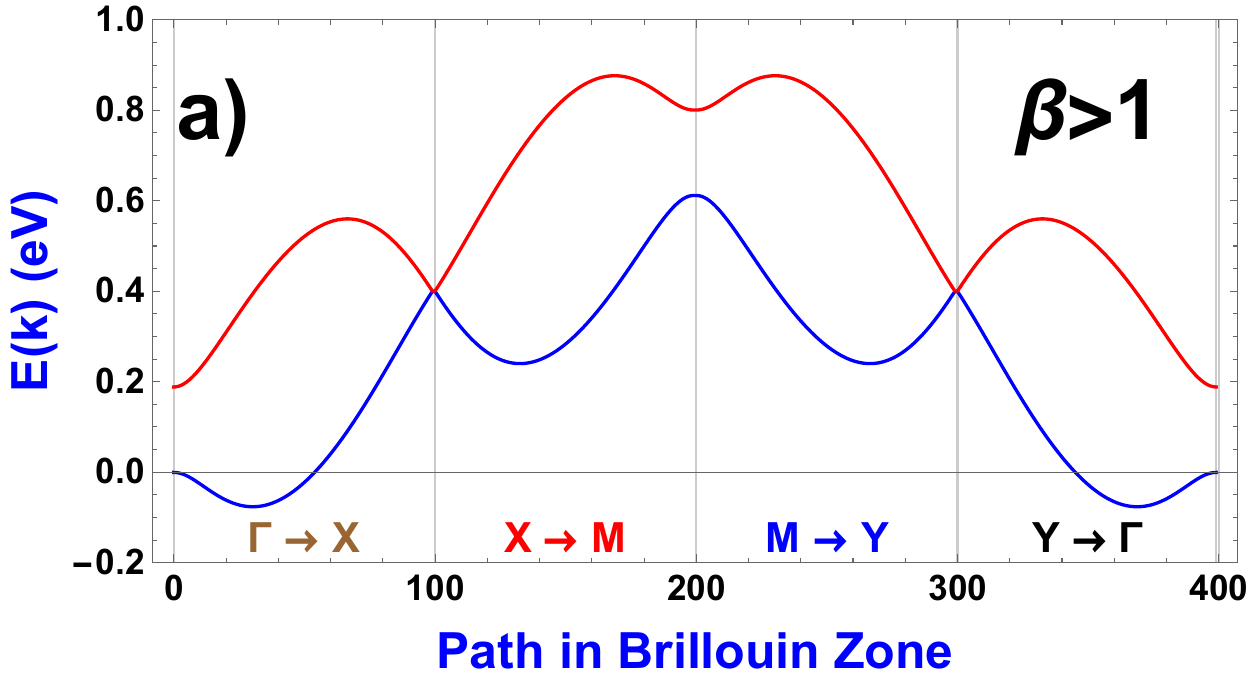}
\includegraphics[width=0.7\linewidth]{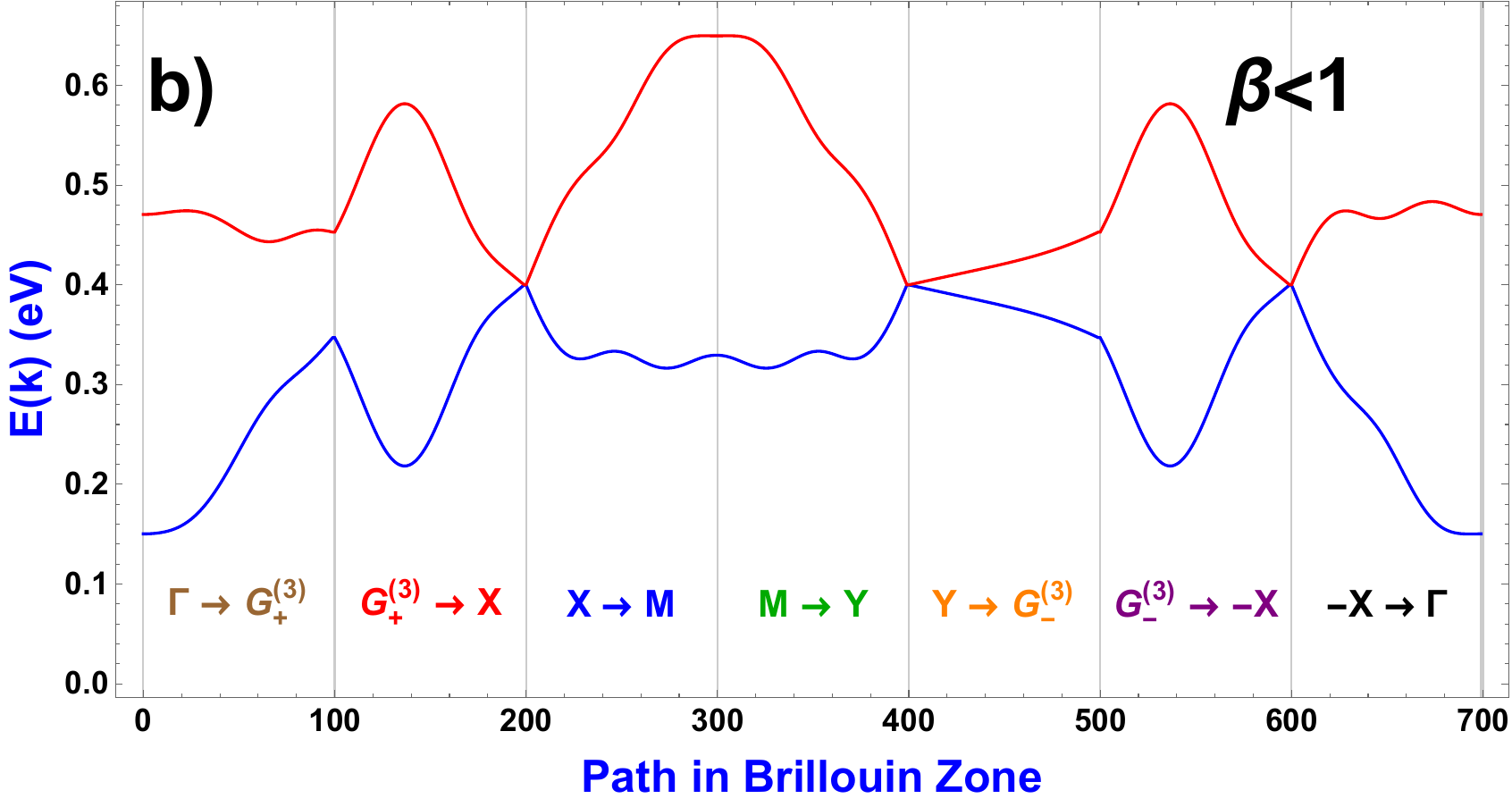}
\includegraphics[width=0.7\linewidth]{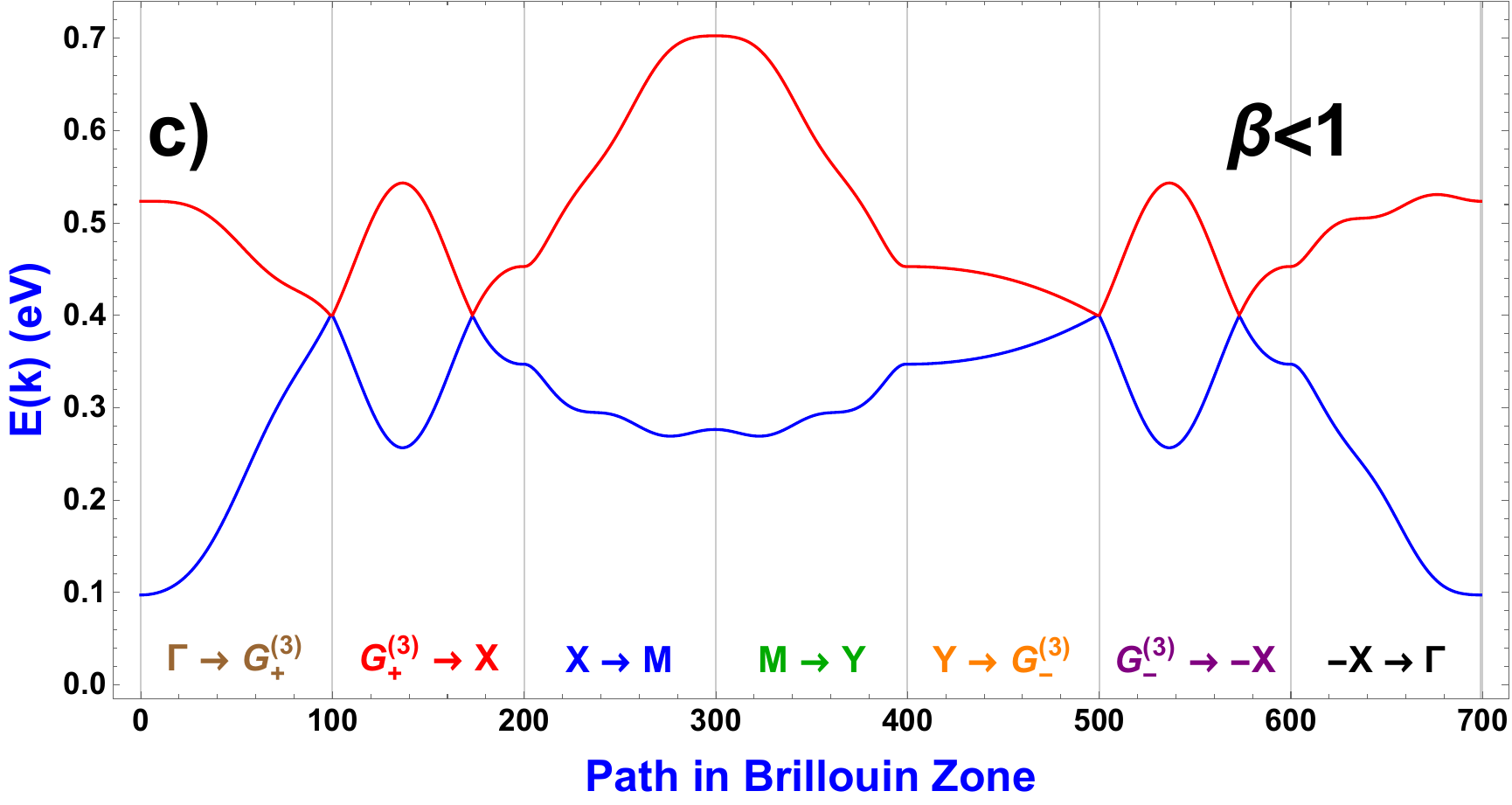}
\caption{Band structure of an irradiated $d_{xy}$-wave altermagnet driven by right-handed circularly polarized light for $(\mathcal{A}_0~\text{nm}^{-1}, M_z~\text{eV})$: (a) $(0.2, 0.047)$, (b) $(0.4, 0.08)$, (c) $(0.4, 0.133)$. The parameter $\beta$ is $2.934 > 1$ for (a) and $0.633 < 1$ for (b) and (c). Consequently, no $G$-family nodes appear in (a), whereas the $G^{(3)}$ points exist in (b) and (c). For case (a), the Brillouin zone path goes through the high-symmetry points, and in agreement with Fig.~\ref{phase_xy}(a), gap closings occur at $X$ and $Y$. For (b) and (c), the path is $\Gamma \rightarrow G^{(3)}_+ \rightarrow X \rightarrow M \rightarrow Y \rightarrow G^{(3)}_- \rightarrow -X \rightarrow \Gamma$, where the light-induced Dirac points $G^{(3)}_\pm$ are located at $(k_x = \pm k_0,\; k_y = \mp \pi/a \pm k_0)$ with $k_0 = a^{-1}\arccos\sqrt{\beta}$. Other parameters: $J = 1$~eV, $\lambda = 0.35$~eV, $\Omega = 2$~eV, $t = 0.1$~eV.}
\label{Band_structure}
\end{figure}

Topological properties of the full-lattice static Hamiltonian represented in Eq.~\ref{hamilk} are examined by calculating the $z$-component of the Jacobian $j_z=(\partial_{k_x}{\vec d}\times\partial_{k_y}{\vec d})_z$, which determines the chirality of each Dirac point. The ${\vec d}$ vector is obtained from the Hamiltonian when it is written as $H_{\text{static}}(\mathbf{k}) = \mathbf{d} \cdot \boldsymbol{\sigma}$. It is straightforward to see that local gap closings occur at the TRIM $\Gamma=(0,0)$, $X=(\pi/a,0)$, $Y=(0,\pi/a)$, $M=(\pi/a,\pi/a)$. The chirality at these points is given by $j_z|_{\Gamma,M} = -j_z|_{X,Y} = (\lambda a)^2$. The mass term for all high-symmetry points is $M_z$. As a result, the Chern number of the lower band for four Dirac points at TRIM reads~\cite{Geometrical}
\begin{equation}
\mathcal{C}=\frac{1}{2}\operatorname{sgn}(M_z) \sum_{\mathbf{k}\in \text{TRIM}}\operatorname{sgn}(j_z[\mathbf{k}])=0.
\label{static_chern}
\end{equation}
Thus, in the static case, independent of the value of $M_z$, the contributions from the high-symmetry points cancel out in the Chern number calculation giving rise to topologicaly trivial altermagnet.  As we will show shortly, the chirality of each gap-closing point and its corresponding mass term change upon application of a CPL which leads to rich topological phases.  Moreover, additional generic Dirac points emerge away from the high-symmetry points.

The real-space tight-binding form of Hamiltonian represented in Eq.~\ref{hamilk} reads as the following
\begin{equation}
\label{TB_Hamiltonian}
\begin{aligned} 
H^{static}=&\sum_{\bf r}^{}[ {\bf c}_{\bf r}^\dagger T_0{\bf c}_{\bf r}+({\bf c}^\dagger_{\bf r+a\hat{x}} T_x{\bf c}_{\bf r}+{\bf c}^\dagger_{\bf r+a\hat{y}} T_y{\bf c}_{\bf r}\\
&{\bf c}^\dagger_{\bf r+a(\hat{x}+\hat{y})} T_{xy}{\bf c}_{\bf r}+{\bf c}^\dagger_{\bf r+a(\hat{x}-\hat{y})} T_{x,-y}{\bf c}_{\bf r}+h.c.)],
\end{aligned}
\end{equation} 
where ${\bf c}_{\bf r}^\dagger$ and ${\bf c}_{\bf r}$ are the creation and annihilation operators of electron at site $\bf{r}$ and 
the onsite and hopping matrices are defined as 
\begin{equation}
\begin{aligned} 
T_0=&4 t \sigma_0+M_z \sigma_z,\,\ T_{xy}=- T_{x,-y}=-\frac{J}{4}\sigma_z,\,\\
T_x=&-t \sigma_0+\frac{\lambda}{2i}\sigma_y,\,\ T_y=-t \sigma_0-\frac{\lambda}{2i}\sigma_x\\
\end{aligned}
\label{static_TB}
\end{equation}
\section{Floquet Hamiltonian}\label{S3}
We use Floquet formalism to provide the evolution of the system at stroboscopic time-scales~\cite{shirley}. The Hamiltonian driving by circularly-polarized light is obtained by means of time-periodic variation of vector potential as ${\textbf A}(t)=A_0(\sin\Omega t,\cos\Omega t)$, where the sign of the light frequency $\Omega$ determines right-handed (positive) and left-handed (negative) circular polarization.  The coupling between light and matter is modeled via the Peierls substitution ${\textbf k}\rightarrow {\textbf k}+\mathcal{A}(t)$ which is applied on the tight-binding Hamiltonian represented in Eq.~\ref{TB_Hamiltonian}, where $\mathcal{A}\equiv {e A/\hbar}$. To find a time-periodic tight-binding Hamiltonian, it is enough to modify the hopping matrices as $T_{\hat{r}} \rightarrow T_{\hat{r}} e^{\mathcal{A}(t).\hat{r}a}$, where $\hat{r}$ denotes to the nearest-neghibour directions on a square lattice along the $x$ and $y$ axes or ($\pm \hat{x} \pm \hat{y}$) diagonal directions\cite{Dabiri2022}. It should be noted that here we assume spacially uniform vector potential in real space.

The Schroedinger equation is explicitly time-periodic, so it is naturally suited for Fourier-transformation in time. Based on the Fourier component of this equation, the Floquet Hamiltonian is defined as   
\begin{equation}
\centering
(H_F)_{m,m'}=H^{(m-m')}-m\Omega\delta_{mm'}
\label{HF}
\end{equation}
where the Fourier components of the time-periodic Hamiltonian are defined as
\begin{equation}
H^{(m)}=1/T \int_{0}^{T} H(t) e^{ i m |\Omega| t} dt.
\end{equation}
The $n^{th}$ Fourier components of the time-periodic Hamiltonian $H^{n}$, is derived as
\begin{equation}
H^{(n)}=T_0 \mathcal{\delta}_{n,0} +\mathcal{J}_n(a\mathcal{A}_0) \mathcal{F}(n)
\label{hf}
\end{equation}
where the second term and so $ \mathcal{F}$ function is valid for $n \ne 0$
$$\mathcal{F}(n)=\sum_{j=1}^8 \mathcal{Q}_j \mathcal{P}_j^n$$
By doing a simple algebra $\mathcal{Q}$ and $\mathcal{P}$ functions read as
\begin{equation}
\label{Hn}
\begin{aligned} 
\mathcal{Q}_1= T_x e^{ik_xa},\mathcal{Q}_3= T_y e^{ik_ya},\mathcal{Q}_2= \mathcal{Q}_1^\dagger,\ \mathcal{Q}_4= \mathcal{Q}_3^\dagger\\
\mathcal{Q}_5=T_{xy} e^{i(k_x+k_y)a}, \mathcal{Q}_7=T_{x,-y} e^{i(k_x-k_y)a},\ \mathcal{Q}_6= \mathcal{Q}_5^\dagger\\
\mathcal{Q}_8= \mathcal{Q}_7^\dagger, \mathcal{P}_1=-1, \mathcal{P}_3=i, \mathcal{P}_2=-\mathcal{P}_1, \mathcal{P}_4=\mathcal{P}_3^*\\
\mathcal{P}_5=\frac{(-1+i)}{\sqrt{2}},\mathcal{P}_6=\frac{(1-i)}{\sqrt{2}}, \mathcal{P}_7=\mathcal{P}_5^*, \mathcal{P}_8=\mathcal{P}_6^*
\end{aligned}
\end{equation} 

An effective Floquet Hamiltonian in the high-frequency regime can be obtained by using perturbation theory which is a series expansion in powers of the inverse frequency~\cite{Bukov,eckart,bw}.
\begin{equation}
H^{\text{eff}}=H^{(0)}+\sum_{n=1}^{\infty}(n \Omega)^{-1}[H^{(-n)},H^{(+n)}]+O(1/(\Omega)^2)
\label{photon_H}
\end{equation}
The resulting effective Hamiltonian is conveniently expressed in the form $H_{\text{eff}}=\textbf{d}^{\text{eff}}.\boldsymbol{\sigma}$, with

\begin{equation}
\begin{aligned}
d_0^{\text{eff}}&= 4 t -2t \mathcal{J}_0\large[ \cos(k_x a)+\cos(k_y a)\large]\\
d_x^{\text{eff}}&=-\mathcal{J}_0 \lambda \sin(k_y a)+\lambda^{\prime} \sin(2 k_x a) \cos(k_y a)\\
d_y^{\text{eff}}&=  \mathcal{J}_0\lambda\sin(k_x a)-\lambda^{\prime} \sin(2 k_y a) \cos(k_x a)\\
d_z^{\text{eff}}&=J \mathcal{J}_0\sin(k_x a)\sin(k_y a)+J^{\prime} \cos(k_x a) \cos(k_y a)+M_z\\
\label{heff}
\end{aligned}
\end{equation}
Here, the correction strength on SOC induced by light is defined as 
\[
\lambda^{\prime}=\frac{2 \lambda J  }{\Omega}\mathcal{S}_{\lambda}(a\mathcal{A}_0),
\]
while a light-induced spin-splitting is given by 
\[
J^{\prime}=\frac{4\lambda^2 }{\Omega}\mathcal{S}_J(a\mathcal{A}_0).
\]
The oscillating $\mathcal{S}$ functions are defined as 
\begin{equation}
\centering
\begin{aligned} 
&\mathcal{S}_{\lambda}(x)=\sum_{m=1}^{\infty} (-1)^{(m+1)} \sin{(m\pi /4)} \mathcal{J}^2_{m}(x)/m,\\
&\mathcal{S}_J(x)=\sum_{m=0}^{\infty} (-1)^m \mathcal{J}^2_{2m+1}(x)/(2m+1).
\end{aligned}
\end{equation}
where $\mathcal{J}_n$ denotes the Bessel Function of the first kind.  The static Hamiltonian (\ref{hamilk}) is recovered by setting $\mathcal{A}_0=0$ in Eq.~\ref{heff}. As an immediate conclusion, the light-induced gap is openned at the high-symmetry points such that it is $2|M_z+J^{\prime}|$ for the $\Gamma$ and $M$ points and $2|M_z-J^{\prime}|$ for the $X$ and $Y$ points. 

The Floquet correction proportional to $\lambda^{\prime}$ represents a spin-orbit interaction generated by CPL. Unlike the conventional SOC which is renormalized by the $\mathcal{J}_0$ factor, this term introduces higher-harmonic momentum dependence over the entire Brillouin zone\cite{Cheraghchi2025}. Consequently, the light field not only opens gaps at the high-symmetry points through the light-induced spin splitting $J^{\prime}$, but also reshapes the global pseudospin texture, enabling the formation of off-symmetry gap-closing points and unconventional topological transitions beyond the massive-Dirac scenario.

The Floquet-induced spin-splitting term  $J^{\prime}\cos(k_x a) \cos(k_y a)$ breaks the pure $d_{xy}$-wave character of the equilibrium altermagnetic spin splitting. As a result, the driven state should be viewed as a Floquet-engineered mixture of d-wave and extended-s-wave spin-splitting channels rather than a pure $d_{xy}$-wave altermagnet\cite{Cheraghchi2025}.

As a short conclusion, the Floquet drive introduces two independent dimensionless ratios. The parameter $\beta \equiv \frac{\mathcal{J}_0 \lambda}{2\lambda'}$  characterizes the competition between the renormalized equilibrium and light-induced SOC's, while $\alpha \equiv \frac{\mathcal{J}_0 J}{J^{\prime}}$ quantifies the competition between the renormalized equilibrium $d_{xy}$-wave spin splitting and the Floquet-induced s-wave spin splitting. Thus, $\beta$ controls the dominant spin-orbit texture and $\alpha$ controls the dominant magnetic symmetry of the driven state.

To quantify how the SOC and out-of-plane magnetization are renormalized by the light illumination, it is useful to examine the low-energy expansion in the effective Hamiltonian in Eq.~\ref{heff} around the $\Gamma$ point in the weak field regime and up to linear order in-k.
\begin{equation}
\begin{aligned}
d_0^{\text{eff}} &= t  a^2  \mathcal{A}_0^2  ,\\
d_x^{\text{eff}} &= (2\lambda^{\prime} k_x-\lambda  k_y) a,\\
d_y^{\text{eff}} &= (\lambda  k_x - 2\lambda^{\prime} k_y) a,\\
d_z^{\text{eff}} &=  J k_x k_y a^2+ J^{\prime} + M_z ,
\label{heff_low}
\end{aligned}
\end{equation}
In the effective out-of-plane spin splitting term $d_z^{\text{eff}}$, the light-induced spin-splitting, $J^{\prime}$, appears as an independent uniform magnetization. By looking at the $d_x^{\text{eff}}$ and $d_y^{\text{eff}}$ terms, it is seen that a linear Dresselhaus SOC is induced by light with the strength $\lambda^{\prime}$ leading to anisotropy in the linear SOC. In the weak field regime, we have $\lambda^{\prime}\simeq \frac{\lambda J (a \mathcal{A}_0)^2}{\sqrt{8}\Omega}$ and $J^{\prime} \simeq  \frac{(\lambda a \mathcal{A}_0)^2}{\Omega}$.  Therefore, the Dresselhaus SOC vanishes when $J=0$, revealing that this is an intrinsic properties arising from the altermagnetism. 

\section{Phase Diagram and Berry Curvature}\label{S4}

The topological properties of the effective Floquet Hamiltonian in Eq.~\ref{heff} are fully determined by the vector $\mathbf{d}=(d_x,d_y,d_z)$, while the scalar term $d_0$ does not affect the topology and Berry curvature. Gap closings occur when $|\mathbf{d}|=0$, and the Chern number can be analytically calculated for the Dirac points emerging at these critical momenta.

\subsection{Dirac-point formulation of the Chern number}\label{S7}

Near each gap-closing point $\mathbf{k}_i$, the Hamiltonian reduces to a massive Dirac form. The contribution of each Dirac point to the Chern number is given by
\begin{equation}
\mathcal{C}_i = \frac{1}{2}\,\operatorname{sgn}(j_z^i)\,\operatorname{sgn}(m_i),
\label{Dirac_formula}
\end{equation}
where $j_z^i$ is the chirality and $m_i$ is the corresponding mass term. The total Chern number is obtained by summing over all Dirac points,
\begin{equation}
\mathcal{C} = \sum_i \mathcal{C}_i.
\end{equation}

For clarity, we separate the contributions into two sectors:
\begin{equation}
\mathcal{C} = \mathcal{C}_{\text{TRIM}} + \mathcal{C}_{G},
\end{equation}
where $\mathcal{C}_{\text{TRIM}}$ arises from high-symmetry points and $\mathcal{C}_G$ from light-induced off-symmetry Dirac points.

\subsection{Gap closings at high-symmetry points}

At the four TRIM, gap closings occur when
\[
M^{\text{Closing}}_z =
\begin{cases}
-J' & \text{at }\Gamma \text{ and } M,\\
+J' & \text{at }X \text{ and } Y.
\end{cases}
\]

The chirality at these points is
\[
j_z(\Gamma,M) = -j_z(X,Y) = (2\lambda' a)^2(\beta^2-1),
\quad
\beta \equiv \frac{\mathcal{J}_0 \lambda}{2\lambda'}.
\]

Each TRIM point contributes
\begin{equation}
\mathcal{C}_{\mathbf{k}} =
\frac{1}{2}\operatorname{sgn}(j_z[\mathbf{k}])\times \operatorname{sgn}(M_z \pm J'),
\end{equation}
where the $+$ ($-$) sign corresponds to $\Gamma,M$ ($X,Y$).

\subsection{Light-induced off-symmetry Dirac points}

For $|\beta|<1$, additional Dirac points emerge away from the TRIM points due to the competition between the renormalized equilibrium and higher-order light-induced SOC terms. Solving $d_x=d_y=0$ yields four families of Dirac points, $G^{(1)}$–$G^{(4)}$, each containing four symmetry-related points in the Brillouin zone (schematic in Fig.~\ref{phase_xy}(b)).

Their locations, gap-closing conditions, and chiralities are summarized in Table~\ref{tab:dirac}. Notably,
\begin{itemize}
\item $G^{(1)}$ and $G^{(2)}$ have positive chirality,
\item $G^{(3)}$ and $G^{(4)}$ have negative chirality.
\end{itemize}

These Dirac points are absent in the static system and are induced purely by the periodic drive.

\subsection{Chern number in different regions}

The phase boundaries shown in Fig.~\ref{phase_xy}(a) are determined by the gap-closing conditions discussed above. The Chern number in each region can be obtained by summing the contributions of all Dirac points according to Eq.~\eqref{Dirac_formula}.
\begin{figure}
\includegraphics[width=\linewidth]{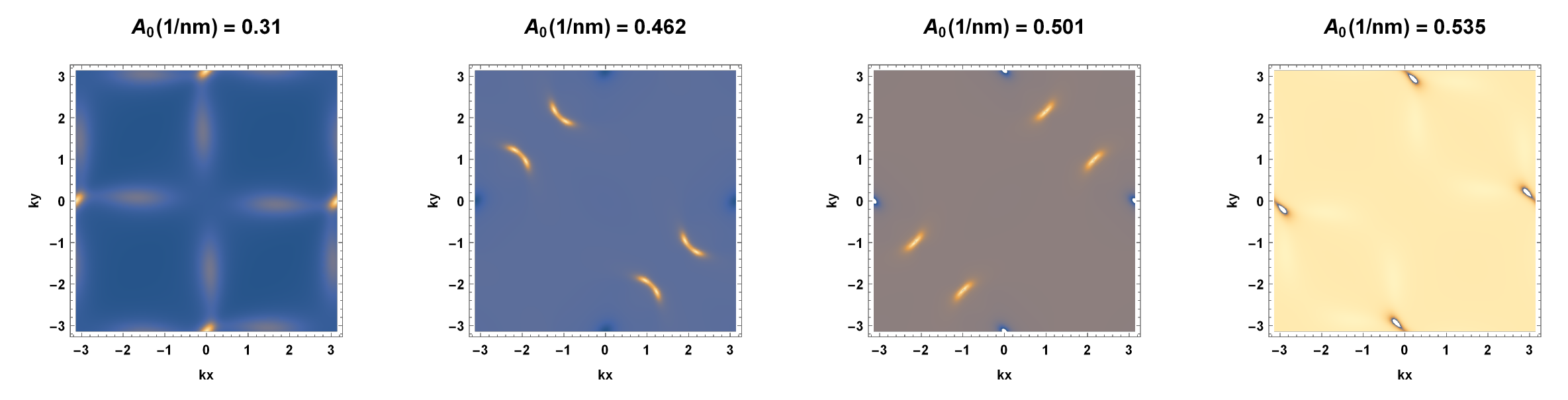} 
\includegraphics[width=\linewidth]{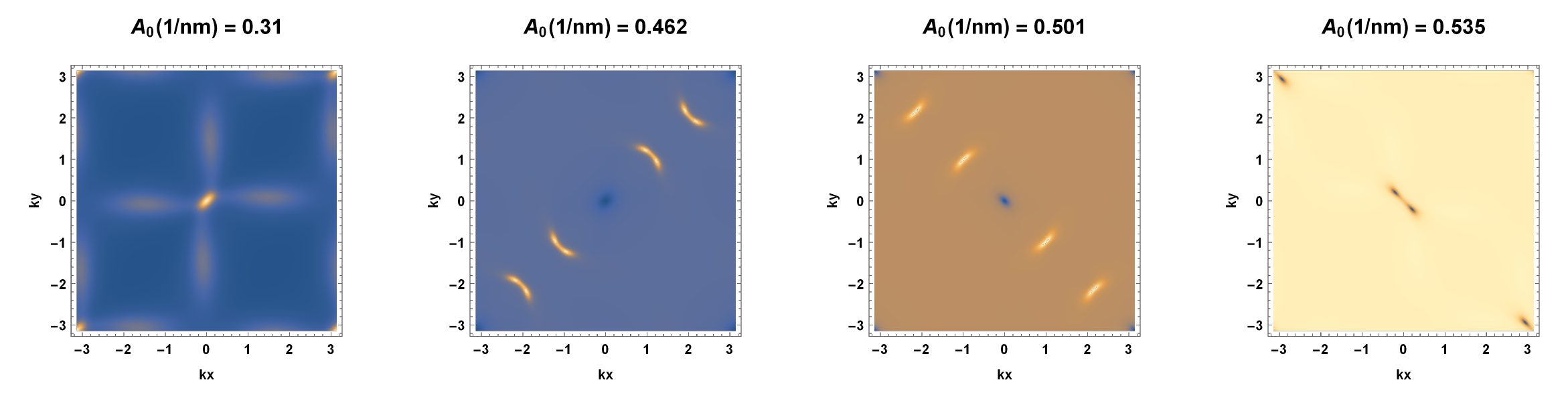}
\caption{Evolution of the Berry curvature of the lower band over the Brillouin zone as a function of light intensity for fixed magnetization. (Top row) For $M_z = +0.05$ eV, the topological charge localization follows the sequence $(X,Y) \rightarrow G^{(3)} \rightarrow G^{(4)} \rightarrow G \rightarrow (X,Y)$. (Bottom row) For $M_z = -0.05$ eV, the sequence is $(\Gamma,M) \rightarrow G^{(1)} \rightarrow G^{(2)} \rightarrow (M,\Gamma)$. Here $G^{(1)}$, $G^{(2)}$, $G^{(3)}$, $G^{(4)}$ denote specific off-symmetry light-induced Dirac points in the Brillouin zone (defined in Table.~\ref{tab:dirac}). The parameters are absoloutly the same as what represented in Fig.~\ref{phase_xy}.}  
\label{Berry_curvature}
\end{figure}
\subsubsection{Regime $|\beta|>1$: TRIM-dominated phases}

In this regime, only the four TRIM points contribute. As an example, consider the yellow region at low intensity regime shown in Fig.~\ref{phase_xy}(a) for the right-handed CPL with $\beta>1$. The contributions are summarized as

\begin{center}
\begin{tabular}{c c c c c}
\toprule
Family &  $ \operatorname{sgn}(j_z)$ & $\operatorname{sgn}(m)$ & Contribution \\
\midrule
$\Gamma, M$ & $+$ & $+$ & $2 \times (1/2)(+1)(+1) = +1$ \\
$X, Y$& $-$ & $-$ & $2 \times (1/2)(-1)(-1) = +1$ \\
\bottomrule
\end{tabular}
\end{center}
The yellow region lies in $M_z \in [-J^{\prime},J^{\prime}]$ (the above of the $\Gamma$/$M$ line and below of the line $X$/$Y$) . Thus, the total Chern number is
\begin{equation}
\mathcal{C} = +2.
\end{equation}
The same calculation can be done for the green region just by exchanging the line $\Gamma$/$M$ with the line $X$/$Y$ giving rise to  $\mathcal{C} = -2.$

To gain comprehensive insight, the band structure derived from the effective Hamiltonian \ref{heff} is plotted in Fig.~\ref{Band_structure}(a) for the region with $\beta>1$, exactly at the gap-closing condition for the $X$ and $Y$ points in the limit of small light intensities. In this regime, the TRIM points dominate the band inversions and consequently play the main role in the Chern number calculation. Notably, an anisotropy is observed in the Dirac cone near the gap-closing points. We will analytically explore this anisotropy in the next section.

\subsubsection{Regime $|\beta|<1$: TRIM and G Points}
As shown in Fig.~\ref{phase_xy}(a), there are yellow regions in intermediate intensities with $|\beta|<1$. In these regions both types of Dirac points (TRIM+G points) contribute in the Chern number. The contribution in the interval $\beta \in [0,1]$ is given as   
\begin{center}
\begin{tabular}{c c c c c}
\toprule
Family & Number & $\operatorname{sgn}(j_z)$ & $\operatorname{sgn}(m)$ & Contribution \\
\midrule
$\Gamma,M$ & 2& - & + & $2 \times (1/2)(-1)(+1) = -1$ \\
$X,Y$&2&+&-&$2 \times (1/2)(+1)(-1) = -1$\\
$G^{(1)}$ & 4 & $+$ & $+$ & $4 \times [+(1/2)] = +2$ \\
$G^{(3)}$ & 4 & $-$ & $-$ & $4 \times [+(1/2)] = +2$ \\
\bottomrule
\end{tabular}
\end{center}
Therefore,
\begin{equation}
\mathcal{C} = +2.
\end{equation}
The same calculation can be done for the green region leading to $\mathcal{C} = -2$.

For intermediate light intensities with $0<\beta<1$, the band structure in Fig.~\ref{Band_structure}(b) is plotted along a path that crosses the TRIM points and four $G^{(3)}$ points. Here, $M_z$ is chosen on the $X$ and $Y$ gap-closing line (the blue solid line in Fig.~\ref{phase_xy}(a)), which lies on the border between the cyan and yellow regions. As seen in Fig.~\ref{Band_structure}(b), the band is closed at the TRIM points but remains gapped at the $G^{(3)}$ points.
 
\subsubsection{Regime $|\beta|<1$: Floquet-induced high-Chern phases}

When $|\beta|<1$, off-symmetry Dirac points appear and contribute to the topology. In particular, in the cyan region of Fig.~\ref{phase_xy}(a), the TRIM contributions cancel, and the Chern number is entirely determined by the $G$ points.

The contributions for interval $\beta \in [0,1]$ are summarized as

\begin{center}
\begin{tabular}{c c c c c}
\toprule
Family & Number & $\operatorname{sgn}(j_z)$ & $\operatorname{sgn}(m)$ & Contribution \\
\midrule
TRIM & 4 & mixed & mixed & $0$ \\
$G^{(1)}$ & 4 & $+$ & $+$ & $4 \times [+(1/2)] = +2$ \\
$G^{(3)}$ & 4 & $-$ & $-$ & $4 \times [+(1/2)] = +2$ \\
\bottomrule
\end{tabular}
\end{center}

Therefore,
\begin{equation}
\mathcal{C} = +4.
\end{equation}

This demonstrates that the high Chern number phase originates from the constructive contribution of light-induced Dirac points, while the TRIM contributions cancel out. The same calculation can be done for the intensity window with $\beta \in [-1,0]$. 

The band structure of the irradiated $d_{xy}$-wave altermagnet driven by right-handed CPL is shown in Fig.~\ref{Band_structure}(c) for parameters where the gap closes at the $G^{(3)}$ points, leading to a high Chern number. Here $0<\beta<1$. Two gap-closing $G^{(3)}$ points appear along the path $\Gamma \rightarrow G^{(3)}_+ \rightarrow X$, and two additional points emerge along $Y \rightarrow G^{(3)}_- \rightarrow -X$, where $-X = (-\pi/a, 0)$ is equivalent to $X$.
\subsection{Physical interpretation}

The phase diagram naturally separates into two regimes:

\begin{itemize}
\item For $|\beta|>1$, the topology is governed solely by TRIM points, yielding conventional phases with $|\mathcal{C}|= 2$.
\item For $|\beta|<1$, the periodic drive generates additional Dirac points away from high-symmetry points, enabling higher Chern numbers such as $|\mathcal{C}|=4$.
\end{itemize}

This mechanism highlights the crucial role of Floquet engineering in generating topological phases beyond those accessible in static systems. All Chern number of the occupied bands reported in the phase diagram (Fig.~\ref{phase_xy}(a)) are numerically confirmed by using the Fukui–Hatsugai–Suzuki discretization scheme \cite{Discrete_BZ}.
\subsection{Berry Curvature Redistibution}
The Chern number is also calculated by utilizing Berry cruvature,
\begin{equation}
\mathcal{C} =
\int_{\mathrm{BZ}}
\frac{d^2\mathbf{k}}{(2\pi)} \,
\Omega_{-}(\mathbf{k}),
\label{eq:sigma_xy_berry}
\end{equation}
where $\Omega_{-}(\mathbf{k})$ is the Berry curvature of the lower (occupied) band and can be computed by using Kubo formula.
\begin{equation}
\Omega_{-}(\mathbf{k}) = 
-2\,\mathrm{Im}
\frac{
\langle -, \mathbf{k} | \hat{v}_x |+, \mathbf{k} \rangle
\langle +, \mathbf{k} | \hat{v}_y | -, \mathbf{k} \rangle
}{
\left(E_{+,\mathbf{k}} - E_{-,\mathbf{k}}\right)^2}.
\label{eq:berry_curvature}
\end{equation}
where $E_{\pm,\mathbf{k}}$ is the upper (lower) band of the effective Floquet Hamiltonian, Eq.~\ref{heff}  and $v_{x}=\frac{1}{\hbar}\frac{\partial H_{\text{eff}}}{\partial k_x}$  is the velocity operator along the x-direction.

Investigation of the Berry curvature evolution by tuning the light amplitude and also $M_z$ helps us to undertand how the phase transition happens and pronounced hot spots emerge around the gapless $G$ points. The evolution of the topological charge as a function the light amplitude is investigated in the Berry curvatures shown in Fig.~\ref{Berry_curvature} for two values of $M_z=\pm 0.05$~\text{eV}. At low intensities, topological charge localizes around the ($X,Y$) points for the value of $M_z=+0.05$~\text{eV}. By enhancing the light amplitude, topological charge is redistributed among the BZ and is accumulated on the $G^{(3)}$ and then $G^{(4)}$ points and finally, it becomes localized around ($X,Y$) points. This sequence of charge localization would be as $(\Gamma,M) \rightarrow G^{(1)} \rightarrow G^{(2)} \rightarrow (M,\Gamma)$ for $M_z=-0.05$~\text{eV}. These sequences of migration from TRIM to G-family points and vice versa, are in agreement with what the phase diagram in Fig.~\ref{phase_xy} (a) and also Table.~\ref{tab:dirac} propose. The emergence of Berry curvature hotspots directly reflects the appearance of Dirac points discussed in Sec.~\ref{S7}. Each hotspot corresponds to a massive Dirac cone and carries a quantized topological charge of $\pm \frac{1}{2}$.

As an immidiate consequence, the topological phase transition is accompanied not only by a gap closing, but also by a qualitative redistribution of Berry curvature in momentum space, with spectral weight transferring from TRIM points to light-induced off-symmetry Dirac nodes.

\subsection{Topological phases in different system parameters}
\begin{figure}
\includegraphics[width= \linewidth]{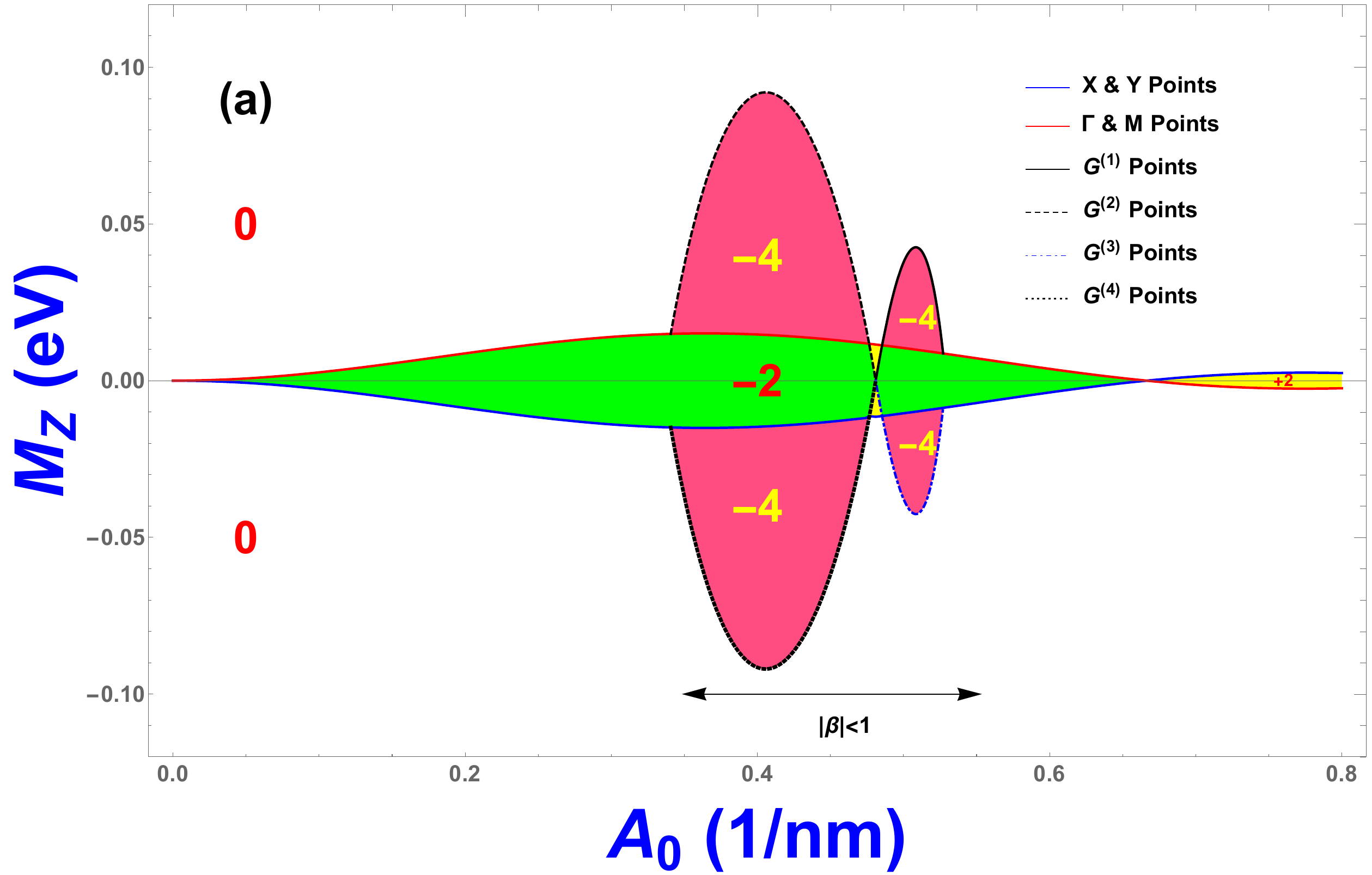} 
\includegraphics[width= \linewidth]{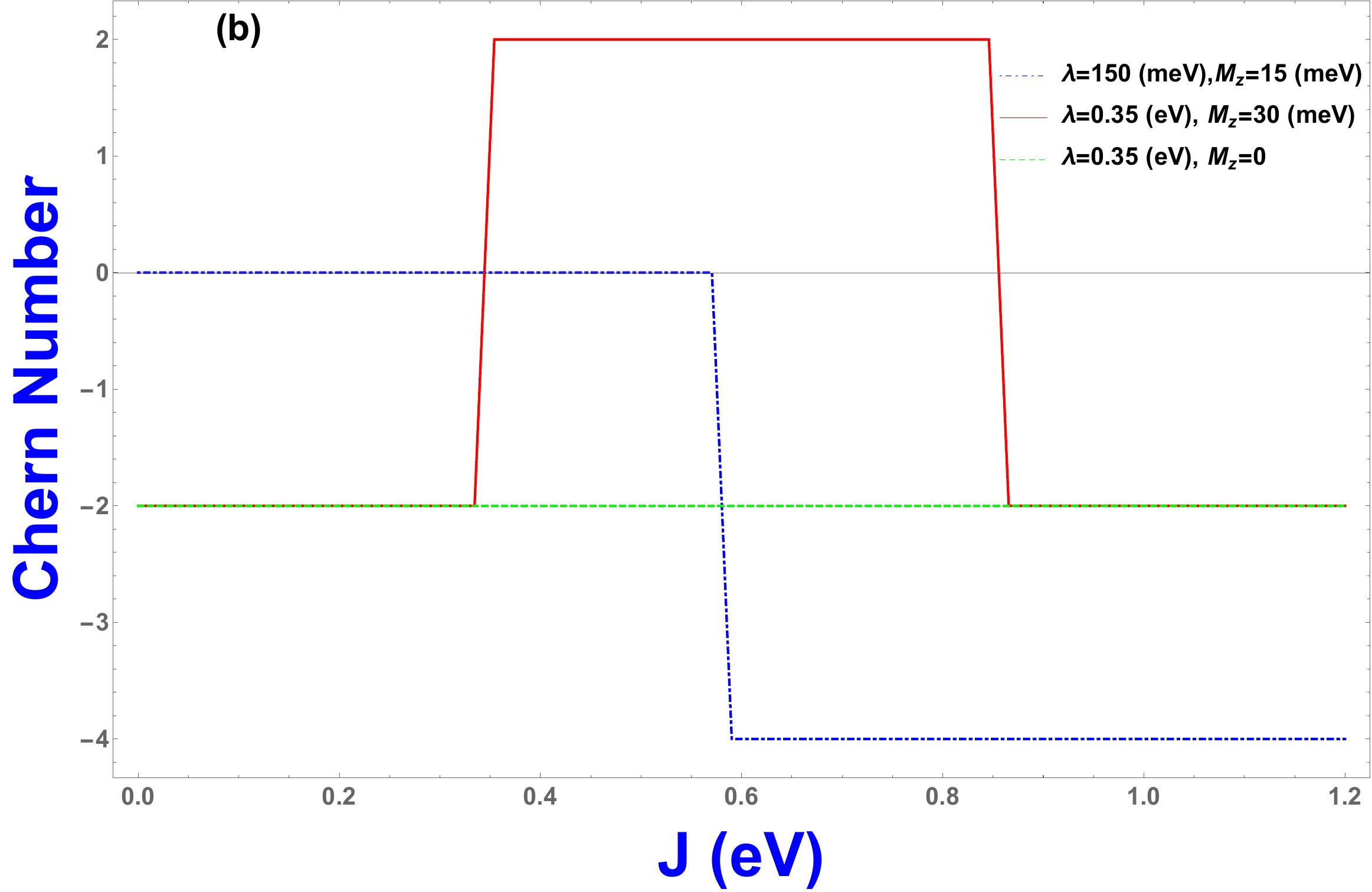} 
\caption{(a) The phase diagram of $d_{xy}$- wave altermagnet irradiated by \textbf{left}-handed circularly polarized light in a high frequency regime and in the presence of out-of-plane magnetization. The parameters are $\lambda=0.15 ~\text{eV}$ and the altermagnetic spin splitting, $J=1.0 ~\text{eV}$. (b) The Chern number of the lower band as a function the altermagnet exchange energy for different system parameters. For this case, the light amplitude is supposed to be $\mathcal{A}_0=0.47~\text{(1/nm)}$. The light frequency is $\Omega=-2~\text{eV}$ for all calculations. } 
\label{left_handed}
\end{figure}

To gain a complementary insight, the phase diagram is re-derived for a smaller SOC energy. As shown in Fig.~\ref{left_handed}(a), the phase diagram is plotted for $\lambda=0.15~\text{eV}$ for the \textbf{left}-handed CPL. Compared to the phase diagram in Fig.~\ref{phase_xy}(a), the QAH phase with high Chern number $|\mathcal{C}|=4$, emerges over a wider range of light intensities and for smaller external magnetization, $M_z$. It should be highlighted that at zero $M_z$, the $d_{xy}$- wave altermagnets irradiated by CPL still have a non-zero Chern number, $|\mathcal{C}|=2$. 

The altermagnet exchange field also leads to variations in Chern number up to $\Delta \mathcal{C}=\pm 4$. Fig.~\ref{left_handed}(b) shows the Chern number of the lower band as a function the altermagnet exchange energy for fixed light intensity $\mathcal{A}_0=0.47~\text{(1/nm)}$ and for different $\lambda$ and $M_z$.

\begin{figure}
\includegraphics[width=0.8\linewidth]{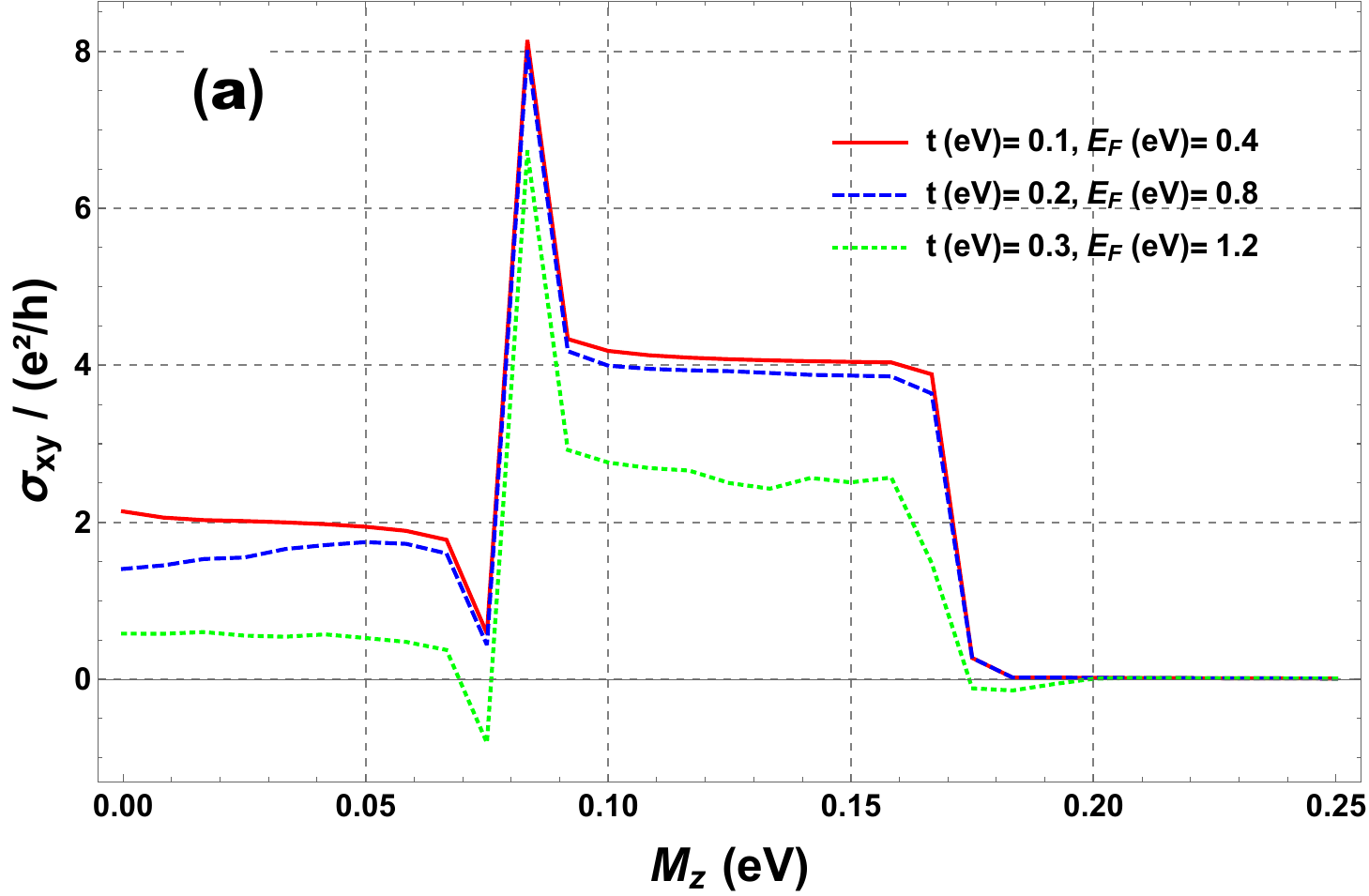} 
\includegraphics[width=0.48\linewidth]{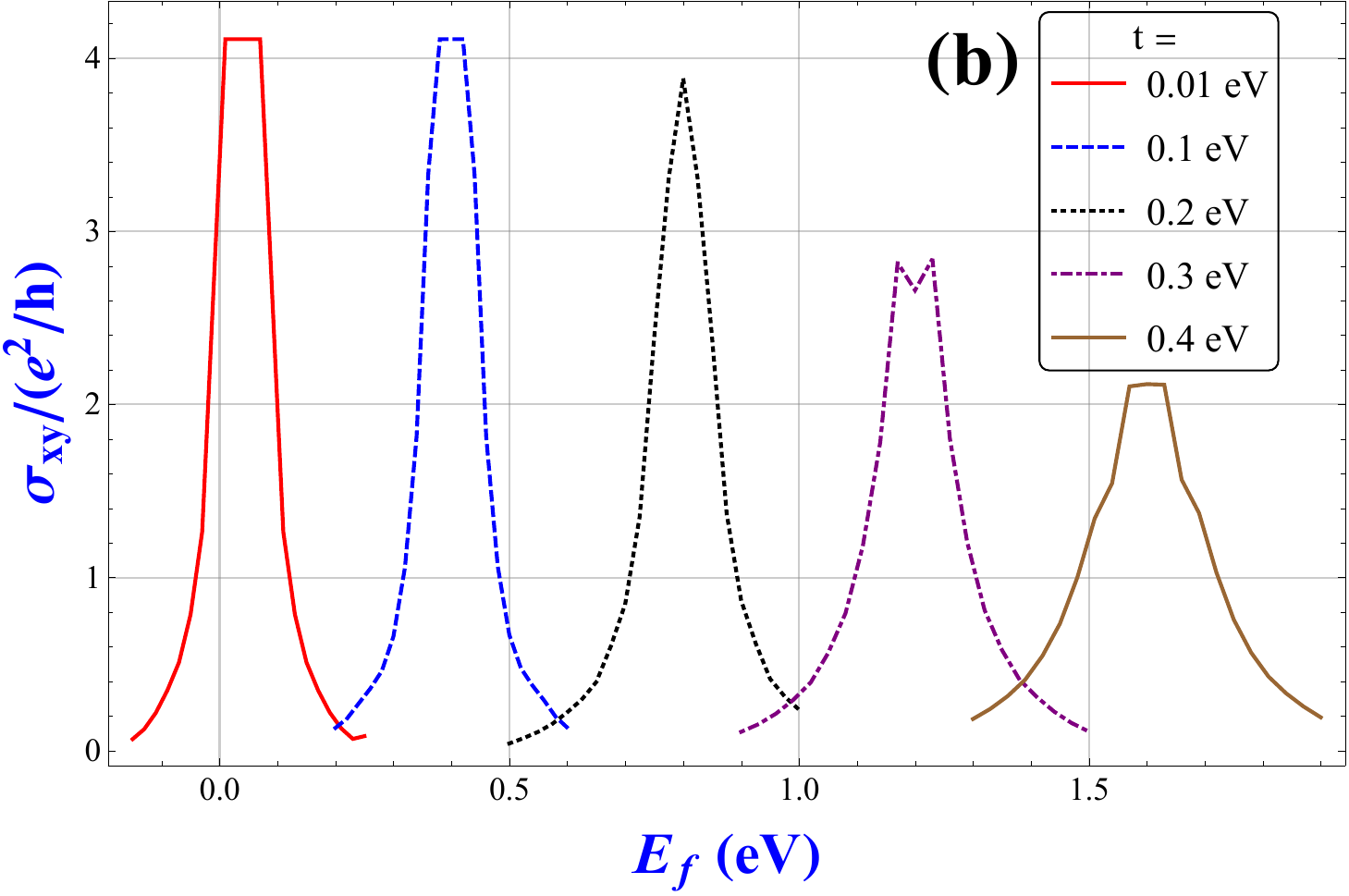}
\includegraphics[width=0.48\linewidth]{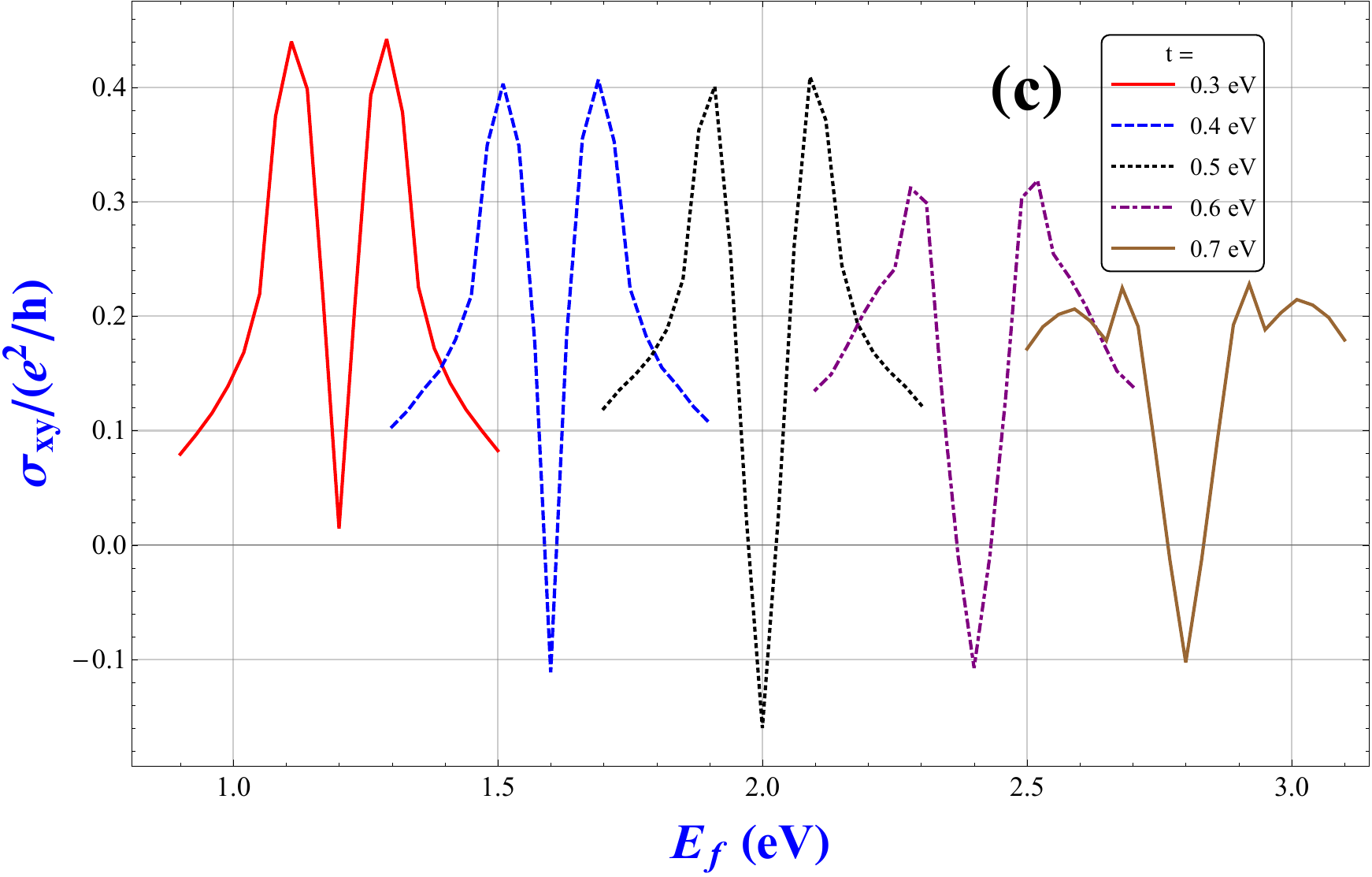}
\includegraphics[width=0.48\linewidth]{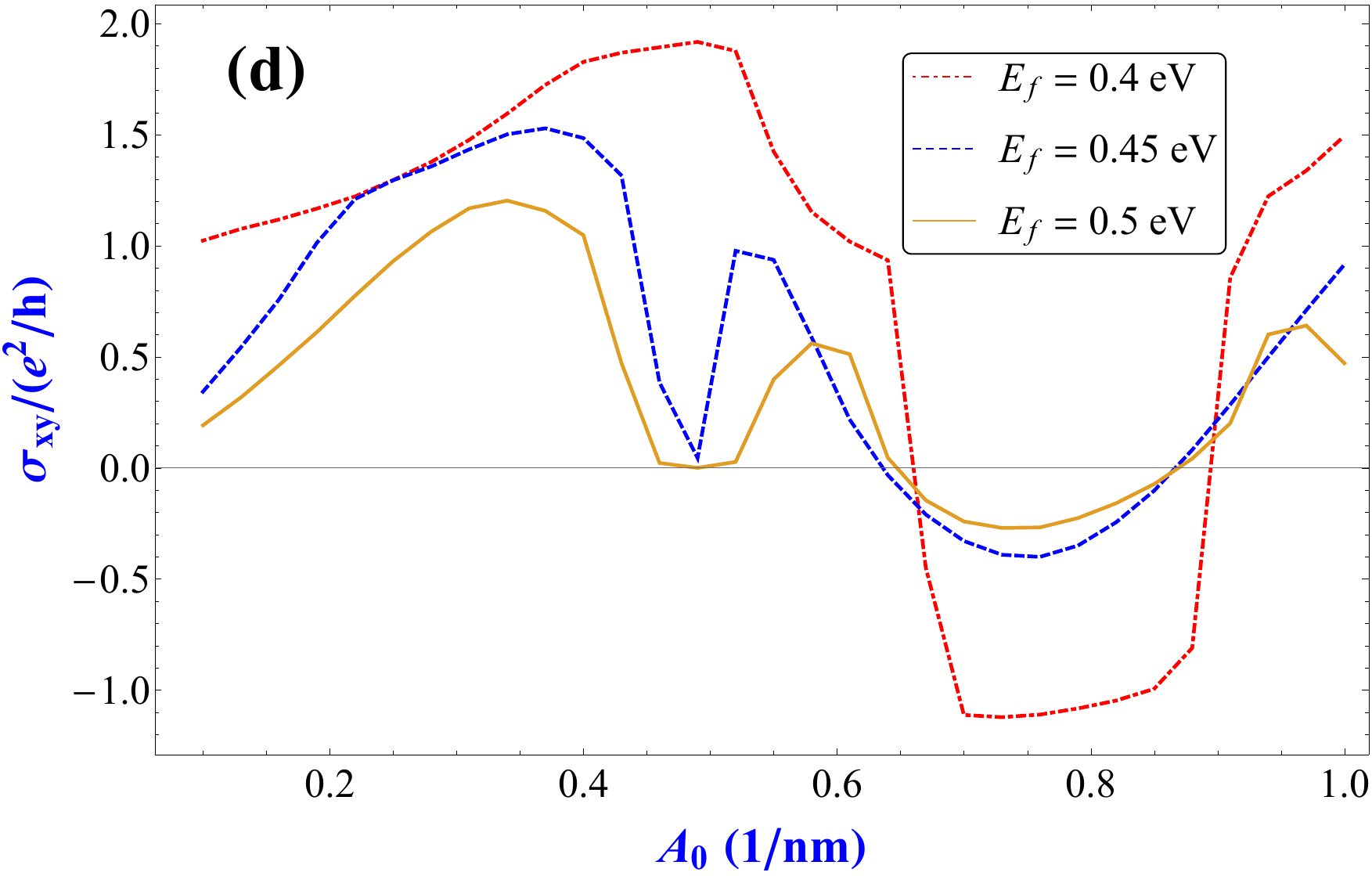} 
\includegraphics[width=0.48\linewidth]{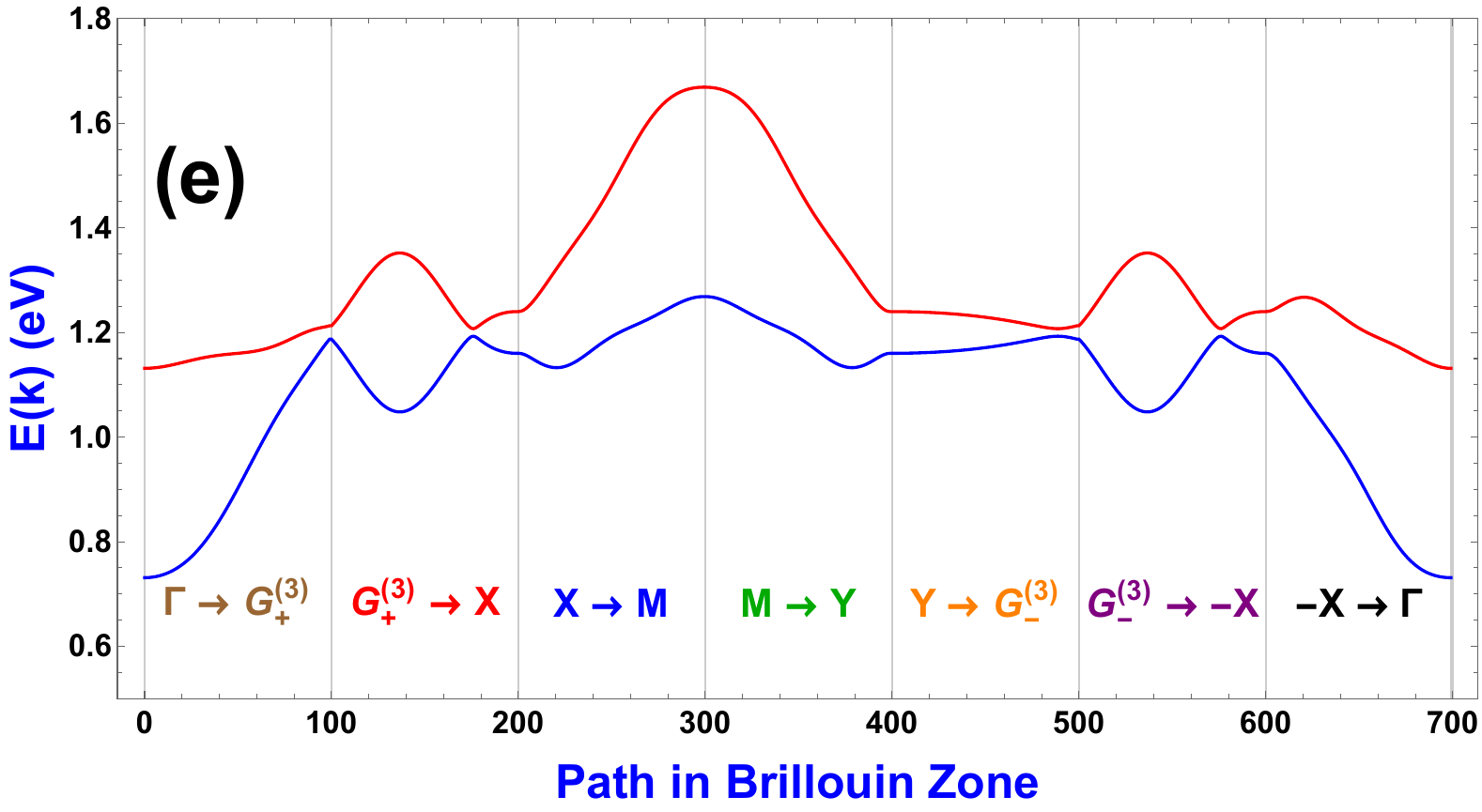} 
\caption{
  Anomalous Hall conductivity $\sigma_{xy}$ (in units of $e^2/h$) tracing the topological phase boundaries of Fig.~\ref{phase_xy}(a).
  \textbf{(a)} $\sigma_{xy}$ vs.\ magnetization $M_z$ at fixed $\mathcal{A}_0 = 0.4$~nm$^{-1}$ for various hoppings $t$.
  \textbf{(b), (c)} $\sigma_{xy}$ vs.\ Fermi energy $E_F$ for different $t$, with $M_z = 0.12$~eV and $M_z = 0.2$~eV, respectively.
  \textbf{(d)} $\sigma_{xy}$ vs.\ light intensity $\mathcal{A}_0$ for several $E_F$, at $t = 0.1$~eV and $M_z = 0$.
  \textbf{(e)} Band structure of the irradiated $d_{xy}$-wave altermagnet with right-handed circularly polarized light at $t = 0.3$~eV and $M_z=0.12$~eV, illustrating gap closure and concentrated Berry curvature at the $G^{(3)}$ gapless points.
  All panels share $J = 1.0$~eV, $\lambda = 0.35$~eV, and $\Omega = 2.0$~eV.
}
\label{AHC}
\end{figure}
\section{Low-energy Hamiltonians and anisotropy}\label{S5}

We analyze the effective Hamiltonian \(H^{\text{eff}} = d^{\text{eff}}_0 + \mathbf{d}^{\text{eff}}\cdot\boldsymbol{\sigma}\) near the gap‑closing points. The kinetic term \(d^{\text{eff}}_0\) does not affect topological or geometric properties. The vector \(\mathbf{d}^{\text{eff}}\) is expanded around a Dirac point at \(\mathbf{k}_0\) as
\[
d^{\text{eff}}_i(\mathbf{k}_0+\mathbf{q}) \approx \mathbf{v}_i\!\cdot\!\mathbf{q} + \mathcal{O}(q^2),
\]
with \(\mathbf{v}_i = \nabla_{\mathbf{k}} d^{\text{eff}}_i|_{\mathbf{k}_0}\) for \(i=x,y,z\). The in‑plane kinetic energy is \(E_{\parallel}^2 = d_x^2+d_y^2 = \mathbf{q}_{\parallel}^T\mathsf{M}\,\mathbf{q}_{\parallel}\) where \(\mathbf{q}_{\parallel}^T=(q_x,q_y)\) is in-plane momentum, \(\mathsf{M} = \mathsf{V}_{\parallel}^T\mathsf{V}_{\parallel}\) is a symmetric positive-definite matrix and \(\mathsf{V}_{\parallel}\) is the \(2\times2\) in-plane velocity matrix for \((d_x,d_y)\). The eigenvalues of \(\mathsf{M}\) give the principal Fermi velocities squared, and their ratio defines the anisotropy of the Dirac cone.

\subsection{High‑symmetry TRIM points}

At the four TRIM points (\(\Gamma, X, Y, M\)), the gap closes when \(M_z = \mp J'\). By symmetry, \(\partial_{k_x}d_z = \partial_{k_y}d_z = 0\), so \(d_z\) is quadratic in \(\mathbf{q}\) (or a constant mass). The linearized effective Hamiltonian reads
\[
H^{\text{eff}}_{\text{TRIM}}(\mathbf{q}_{\parallel}) = 
\begin{pmatrix}
v_{xx} & v_{xy} \\
v_{yx} & v_{yy}
\end{pmatrix}
\begin{pmatrix} q_x \\ q_y \end{pmatrix}
\cdot \boldsymbol{\sigma}_{\parallel} + \mathcal{O}(q^2)\,\sigma_z,
\]
with the velocity matrices
\[
\mathsf{V}_{\parallel}|_{\Gamma} = a\begin{pmatrix}
2\lambda' & -\mathcal{J}_0\lambda \\
\mathcal{J}_0\lambda & -2\lambda'
\end{pmatrix},\qquad
\mathsf{V}_{\parallel}|_{X} = a\begin{pmatrix}
2\lambda' & -\mathcal{J}_0\lambda \\
-\mathcal{J}_0\lambda & 2\lambda'
\end{pmatrix}
\]
where, \( \mathsf{V}_{\parallel} |_{M}=-\mathsf{V}_{\parallel} |_{\Gamma}\) and \( \mathsf{V}_{\parallel} |_{Y}=-\mathsf{V}_{\parallel} |_{X}\). The effective metric for the in-plane kinetic energy is obtained as
\[
\mathsf{M}_{\text{TRIM}} = \mathsf{V}_{\parallel}^T\mathsf{V}_{\parallel}
= 4\lambda'^2 a^2 \begin{pmatrix}
1+\beta^2 & -2\beta \\
-2\beta & 1+\beta^2
\end{pmatrix},
\]
with eigenvalues \(4\lambda'^2 a^2(1\pm\beta)^2\). Hence the anisotropy ratio (max/min velocity) is
\[
\eta_{\text{TRIM}} = \frac{1+|\beta|}{|1-|\beta||}.
\]
For \(|\beta|<1\), the cone is anisotropic (except at \(\beta=0\) where it is isotropic); for \(|\beta|>1\) the ratio remains larger than 1. In anisotropic Dirac cones, the iso-energy contours are ellipses rather than circles. Indeed, the metric \(\mathsf{M}\) encodes the directional dependence of the Dirac cone's slope.  The Berry curvature near a TRIM point is well described by the massive Dirac formula \(\Omega_z = \frac12\, m j_z/(m^2+\mathbf{q}_{\parallel}^T\mathsf{M}\,\mathbf{q}_{\parallel})^{3/2}\) with \(j_z = \det\mathsf{V}_{\parallel}\). Here, $m|_{\text{TRIM}}$ is the mass term at the TRIM.

\subsection{Light‑induced off‑symmetry \(G\) points}

For \(|\beta|<1\), additional Dirac points appear away from the TRIM (the \(G\) families). Their locations are given in Table~\ref{tab:dirac}. At these points, \(\partial_{k_x}d_z\) and \(\partial_{k_y}d_z\) are {\it nonzero} (see Table~\ref{tab:velocities}). Therefore \(d_z\) is linear in \(\mathbf{q}\) at leading order. The low‑energy Hamiltonian takes the form
\[
H^{\text{eff}}_{G}(\mathbf{q}_{\parallel}) = \begin{pmatrix}
v_{xx} & v_{xy} \\
v_{yx} & v_{yy}
\end{pmatrix}
\begin{pmatrix} q_x \\ q_y \end{pmatrix}
\cdot \boldsymbol{\sigma}_{\parallel} + \bigl(v_{zx}q_x+v_{zy}q_y\bigr)\sigma_z.
\]
This is a generic anisotropic 2D Dirac point with all Pauli components linear in momentum. On the other words, this is a fully tilted anisotropic two-band linear touching described by a three velocity matrices leading to a 3-component pseudospin texture. Although the resulting structure can be made formally analogous to an anisotropic Weyl-like form, the system does not admit a true Weyl classification due to the lack of a third independent momentum direction required for a topological monopole structure in momentum space. Instead, the node should be understood as a generalized anisotropic Dirac point with a rank-3 velocity tensor.
The in‑plane metric \(\mathsf{M}_{\parallel} = \mathsf{V}_{\parallel}^T\mathsf{V}_{\parallel}\) still governs the dispersion of \(d_x,d_y\), while the third component  \(d_z = \mathbf{v}_z\cdot\mathbf{q}_{\parallel}\) contributes on equal footing as an additional linear kinetic channel. This enhanced pseudospin-momentum entanglement is expected to modify Berry curvature distribution near the node and may have implications for nonlinear and optical response functions.

\subsubsection{Berry curvature decomposition}

The Berry curvature of the lower band for a two‑band model is
\begin{equation}
\Omega_z(\mathbf{q}_{\parallel}) = \frac{1}{2}\,\frac{\mathbf{d}\cdot(\partial_{q_x}\mathbf{d}\times\partial_{q_y}\mathbf{d})}{|\mathbf{d}|^{3}}.
\label{Berry_two_band}
\end{equation}
Writing \(j_i = (\partial_{q_x}\mathbf{d}\times\partial_{q_y}\mathbf{d})_i\) for \(i=x,y,z\), we have
\[
\Omega_z = \Omega_{\text{Dirac}}+\Omega_{\text{geom}}.
\]
where \(\Omega_{\text{Dirac}}=d_z j_z /(2|\mathbf{d}|^{3}) \) and \(\Omega_{\text{geom}}= (d_x j_x + d_y j_y)/(2|\mathbf{d}|^{3}) \). The term \(d_z j_z\) is the usual “Dirac” contribution (present even when \(d_z\) is constant or quadratic), while \(d_x j_x + d_y j_y\) is a "geometric" contribution arising from the in‑plane components of the Jacobian. For TRIM points, symmetry forces \(j_x = j_y = 0\); thus \( \Omega_{\text{geom}}^{\text{TRIM}}=0\) and only the Dirac part survives. For \(G\) points, both \(j_x\) and \(j_y\) are nonzero (see Table~\ref{tab:velocities}), leading to an additional geometric contribution that modifies the Berry curvature distribution, \( \Omega_{\text{geom}}^{\text{G-Points}}\neq 0\). As a result, the Berry curvature near a \(G\) point must be computed from the full vector product and is not captured by the simplified formula used for TRIM.

\subsubsection{Velocity matrices and anisotropy}

For concreteness, we list the velocity matrices for each family in Table.~\ref{tab:velocities} using the parameter \(\gamma = |\beta|\) and the sign factor \(s = \operatorname{sign}\large[\sin (k_x a)\cos (k_x a)\large]\).

\begin{table*}[htbp]
\centering
\caption{Low‑energy velocity matrices at \(G\) points. \(\Lambda = 2\lambda'a\sqrt{\gamma}\), and \(\gamma = |\beta|\).}
\begin{tabular*}{\linewidth}{@{\extracolsep{\fill}} c c c @{}}
\toprule
Family & \(\mathsf{V}_{\parallel}\) & \(\mathbf{v}_z = (\partial_{k_x}d_z,\partial_{k_y}d_z)\) \\
\midrule
\(G^{(1)}\) & \(\Lambda\begin{pmatrix}2\gamma-1&-1\\1&-(2\gamma-1)\end{pmatrix}\) & \(a s (J\mathcal{J}_0-J')\sqrt{\gamma(1-\gamma)}\,(1,1)\) \\
\(G^{(2)}\) & \(\Lambda\begin{pmatrix}2\gamma-1&1\\-1&-(2\gamma-1)\end{pmatrix}\) & \(a s (J\mathcal{J}_0+J')\sqrt{\gamma(1-\gamma)}\,(-1,1)\) \\
\(G^{(3)}\) & \(\Lambda\begin{pmatrix}-(2\gamma-1)&1\\1&-(2\gamma-1)\end{pmatrix}\) & \(a s (J'-J\mathcal{J}_0)\sqrt{\gamma(1-\gamma)}\,(1,1)\) \\
\(G^{(4)}\) & \(\Lambda\begin{pmatrix}2\gamma-1&1\\-1&2\gamma-1\end{pmatrix}\) & \(a s (J\mathcal{J}_0+J')\sqrt{\gamma(1-\gamma)}\,(1,-1)\) \\
\bottomrule
\end{tabular*}\label{tab:velocities}
\end{table*}

The eigenvalues of \(\mathsf{M}_{\parallel} = \mathsf{V}_{\parallel}^T\mathsf{V}_{\parallel}\) for all \(G\) families are identical: \( \Lambda^2\bigl(1\pm|2\gamma-1|\bigr)^2\), where  \(\Lambda = 2\lambda'a\sqrt{\gamma}\). Thus the in‑plane anisotropy ratio is
\[
\eta_G(\gamma) = \frac{1+|2\gamma-1|}{1-|2\gamma-1|},\qquad 0\le\gamma\le1,
\]
which equals 1 at \(\gamma=1/2\) (isotropic cone) and diverges as \(\gamma\to0\) or \(\gamma\to1\) (semi‑Dirac limit).

Let us compare in-plane anisotropy around TRIM and G-points. Taking \(\beta = 0.3\) (so \(\gamma = 0.3\) for the \(G\) family) results in \(\eta_{\text{TRIM}} \approx 1.857,\qquad
\eta_G  \approx 2.333\). Thus the \(G\) point is more anisotropic than the TRIM point for this \(\beta\). At \(\beta=0.5\), \(\eta_{\text{TRIM}}=3\) (strongly anisotropic) while \(\eta_G=1\) (isotropic). This illustrates that the two types of Dirac points can have very different anisotropy responses as the light amplitude varies.

\section{Signatures in anomalous Hall conductivity for topology}\label{S6}

Depending on the strength of the kinetic energy, the system cannot be classified as a conventional topological insulator across the entire hopping regime. In the strong-hopping limit, it instead realizes a topological metallic or semimetallic phase characterized by locally gapped avoided band crossings. In this regime, topology is not encoded in a global bulk gap invariant but rather emerges from Berry curvature concentrated near these avoided crossings. 

To experimentally verify the phase diagram proposed in Fig.~\ref{phase_xy}(a), a particularly suitable observable with sharp topological signatures is the AHC. In metallic systems, the AHC is generally not quantized; however, it remains a direct probe of the Berry curvature distribution in momentum space, especially its accumulation near avoided crossings and local gap openings, thereby reflecting the underlying topology~\cite{Haldane_2004}. 

At finite temperature, the intrinsic AHC, $\sigma_{xy}$ is given by
\[
\sigma_{xy} = \frac{e^{2}}{\hbar} \sum_{n} \int_{\mathrm{BZ}} \frac{d^{2}k}{(2\pi)^{2}} \; \Omega_{n,z}(\mathbf{k}) \; f(\varepsilon_{n\mathbf{k}}),
\]
where $f(\varepsilon) = \bigl(e^{(\varepsilon-E_F)/k_{B}T}+1\bigr)^{-1}$ is the Fermi–Dirac distribution and $E_F$ denotes the Fermi energy. In a fully gapped two-band insulator, the Brillouin-zone integral of $\Omega_{-,z}$ yields a quantized Chern number. In contrast, in metallic systems the AHC is a Fermi-level property such that its variation with the Fermi energy is governed primarily by Berry curvature in the vicinity of the Fermi level,~\cite{Haldane_2004}.

For a generic two-band model, the Berry curvature satisfies $\Omega_{+,z} = -\Omega_{-,z}$, and the Hall conductivity can be rewritten as
\[
\sigma_{xy} = -\frac{e^{2}}{\hbar} \int \frac{d^{2}k}{(2\pi)^{2}} \; \Omega_{-,z}(\mathbf{k}) \; \bigl[ f(\varepsilon_{+,\mathbf{k}}) - f(\varepsilon_{-,\mathbf{k}}) \bigr].
\]
This form makes explicit that only states in the vicinity of the Fermi energy—where the occupation difference between bands is nonzero—contribute effectively to the AHC. By tuning the Fermi energy, one effectively scans through regions of the band structure weighted by Berry curvature. Consequently, sign reversals or sharp peaks in $\sigma_{xy}$ signal that the Fermi level intersects momentum-space regions with strongly localized Berry curvature, typically associated with local band inversions or avoided crossings.

The phase diagram in Fig.~\ref{phase_xy}(a) can therefore be experimentally accessed through measurements of the AHC as a function of out-of-plane magnetization, as shown in Fig.~\ref{AHC}(a) for different hopping energies. These results are obtained using the Kubo formula as expressed in Eq.~\ref{eq:berry_curvature}. The fixed light intensity $\mathcal{A}_0 = 0.4~\mathrm{nm}^{-1}$ corresponds to $\beta = 0.63 < 1$, placing the system in a regime where the $G$-family of symmetry-related gapless points emerges alongside high-symmetry avoided crossings.

As shown in Fig.~\ref{AHC}(a), the evolution of the AHC closely follows the phase structure in Fig.~\ref{phase_xy}(a). In particular, the chosen parameters exhibit a sequence of topological phase transitions characterized by changes in the Chern number $\mathcal{C} = +2 \rightarrow +4 \rightarrow 0$. The sharp variations in the AHC at the transition point ( $\mathcal{C} = +2 \rightarrow +4$) originate from the redistribution of Berry curvature, which becomes strongly localized at the gap-closing points, shifting between the $X$ and $Y$ points and the $G^{(3)}$ points.

Although the AHC is not quantized in metallic regimes and is gradually suppressed by increasing kinetic energy, it still exhibits pronounced kinks and sharp variations that serve as clear signatures of topological phase transitions. A particularly direct probe is the dependence of the AHC on the Fermi energy, shown in Fig.~\ref{AHC}(b,c). For $M_z = 0.12~\mathrm{eV}$, Fig.~\ref{AHC}(b) shows sharp peaks in $\sigma_{xy}$ when the Fermi level passes through local gaps where Berry curvature is strongly concentrated. These features remain robust even as the hopping energies increases and the global gap closes.

In the weak-hopping limit, e.g., $t = 0.01~\mathrm{eV}$, a well-defined plateau of the AHC appears inside the global bulk gap. In contrast, for $M_z = 0.2~\mathrm{eV}$, as shown in Fig.~\ref{AHC}(c), a pronounced dip emerges in the AHC at the Fermi level, consistent with a trivial phase with vanishing Chern number.

Figure~\ref{AHC}(d) presents the dependence of the AHC on light intensity for $M_z = 0$ and various Fermi energies. An enhancement of $\sigma_{xy}$ at intermediate $\mathcal{A}_0$ corresponds to the region in the phase diagram of Fig.~\ref{phase_xy}(a) with Chern number $+2$. At stronger light intensities, a sign reversal in the AHC signals entry into the $-2$ Chern phase. By tuning the Fermi energy, the relative contributions of upper and lower bands to the Berry curvature integral are controlled, and energies with strongly localized Berry curvature exhibit enhanced absolute values of $\sigma_{xy}$.

Finally, Fig.~\ref{phase_xy}(d) shows the band structure of the irradiated $d_{xy}$-wave altermagnet under right-handed CPL at $t = 0.3~\mathrm{eV}$. Gap closing occurs at the $G^{(3)}$ points, where Berry curvature hot spots are formed. However, as shown in Fig.~\ref{phase_xy}(d), small residual mass terms persist at these points. Thus, although the system exhibits global gap closing, the relevant topological features are governed by local avoided crossings. As demonstrated in Figs.~\ref{AHC}(a–d), these Berry curvature hot spots produce robust and experimentally accessible signatures in the AHC.

\section{Conclusion}\label{S7}

In this work, we studied Floquet topological phases in two-dimensional $d_{xy}$-wave altermagnets driven by off-resonant circularly polarized light and subject to an out-of-plane magnetization, $M_z$, induced via extrinsic exchange coupling from a proximate ferromagnet. The light–matter coupling is incorporated via the Peierls substitution, leading to a lattice Floquet Hamiltonian characterized by a dimensionless driving parameter $\beta$. Depending on $\beta$, two distinct classes of gap-closing points emerge.

For $|\beta|>1$, Berry curvature is predominantly concentrated near TRIM Dirac points, resulting in Chern phases with $|\mathcal{C}|=2$. In contrast, for $|\beta|<1$, the periodic drive induces additional off-symmetry gapless points forming four $G$-families, $G^{(1)}$–$G^{(4)}$, which support higher topological phases up to $|\mathcal{C}|=4$. We systematically derived the locations, gap-closing conditions, and chiralities of these $G$ points. The resulting phase diagram in the $(M_z,\mathcal{A}_0)$ plane is thus separated into regimes governed by $|\beta|>1$ and $|\beta|<1$, with all Chern numbers obtained analytically from Berry curvature contributions of individual gap-closing points.

The low-energy analysis shows that TRIM Dirac points exhibit anisotropic massive Dirac dispersions, where the anisotropy ratio depends on $\beta$. Their Berry curvature follows the standard anisotropic massive Dirac form. In contrast, the light-induced $G$ points correspond to additional band-touching solutions characterized by a fully linear expansion of all components of $\textbf{d}(\textbf{k})$. The resulting effective Hamiltonian takes the form of a general anisotropic two-band linear crossing with a rank-3 velocity tensor, where all Pauli components acquire leading-order momentum dependence with enhanced pseudospin–momentum coupling. The term $d_z$ acts as an additional kinetic channel rather than a momentum-independent mass. Consequently, the Berry curvature cannot be described by the standard massive Dirac form. The velocity tensors and anisotropy ratios of the $G$ points were derived explicitly, showing a strong $\beta$ dependence. 

Finally, we discuss experimental signatures of the proposed phase diagram. In the strong-hopping regime, the system becomes a topological metal or semimetal with no global bulk gap. Nevertheless, topology persists through Berry curvature accumulation near locally gapped avoided crossings. In this regime, the anomalous Hall conductivity provides a direct probe of the underlying Berry curvature distribution. Although generally non-quantized, the AHC exhibits sharp kinks, sign changes, and strong variations when the Fermi level intersects regions of concentrated Berry curvature, thereby providing clear signatures of topological transitions.

\section*{Data Availability}
The data that support the findings of this study are available from the authors upon reasonable request.

\end{document}